\documentclass[%
 reprint,
 amsmath,amssymb,
 aps,
]{revtex4-2}

\usepackage{graphicx}
\usepackage{dcolumn}
\usepackage{bm}
\usepackage{hyperref}
\usepackage{xcolor}
\usepackage{braket}
\usepackage{subfigure}
\usepackage{amsmath}
\usepackage{soul}
\usepackage{ulem}
\usepackage{cancel}

\begin{document}

\preprint{APS/123-QED}

\title{Analog Quantum Simulation of Dirac Hamiltonians in Circuit QED Using Rabi Driven Qubits}

\author{Gal Gumpel$^1$}
 \thanks{These authors contributed equally to this work.}
\author{Jiwon Kang$^2$}
 \thanks{These authors contributed equally to this work.}
\author{Eliya Blumenthal$^1$}
\author{Aron Klevansky$^1$}
\author{Eunseong Kim$^2$}
\author{Shay Hacohen-Gourgy$^1$}
 \email{shayhh@physics.technion.ac.il}
\affiliation{$^1$Department of Physics, Technion - Israel Institute of Technology, Haifa 32000, Israel}
\affiliation{$^2$Department of Physics, Korea Advanced Institute of Science and Technology (KAIST), Republic of Korea}

\date{\today}

\begin{abstract}
Quantum simulators hold promise for solving many intractable problems. However, a major challenge in quantum simulation, and quantum computation in general, is to solve problems with limited physical hardware. Currently, this challenge is tackled by designing dedicated devices for specific models, thereby allowing to reduce control requirements and simplify the construction. Here, we suggest a new method for quantum simulation in circuit QED, that provides versatility in model design and complete control over its parameters with minimal hardware requirements. We show how these features manifest through examples of quantum simulation of Dirac dynamics, which is relevant to the study of both high-energy physics and 2D materials. We conclude by discussing the advantages and limitations of the proposed method.
\end{abstract}

\maketitle

\section{\label{sec:intro}Introduction}

The challenge of simulating quantum systems efficiently has long been recognized as a fundamental limitation of classical computation, as the description of nature inherently requires quantum simulations and computations \cite{Feynman1982}. This realization has motivated the development of quantum simulators, which are controllable quantum systems engineered to emulate the properties and dynamics of less-accessible quantum systems \cite{Buluta2009}. By leveraging quantum principles, these simulators enable the exploration of phenomena that are computationally intractable with classical approaches. 

Quantum simulators can be broadly categorized into two types: analog simulators \cite{Friedenauer2008, Lamata2007}, which emulate the target system’s Hamiltonian directly, and digital simulators \cite{PhysRevX.7.031023, doi:10.1126/science.1113479}, which decompose its dynamics through discrete quantum gate operations. Analog simulators are compact and efficient but offer limited flexibility, whereas digital simulators provide more sophisticated control at the cost of increased resource requirements \cite{RevModPhys.86.153}.

These approaches have been successfully implemented across various experimental platforms, ultracold quantum gases \cite{Bloch2012}, trapped ions \cite{Blatt2012}, photonic systems \cite{Aspuru-Guzik2012}, quantum dots \cite{Hensgens2017}, and superconducting circuits \cite{Houck2012,Lamata_2018}. Despite these advances, the scalability and stability of quantum simulators remains a significant challenge. Non-ideal experimental constraints, such as decoherence, limited qubit lifetimes, and gate fidelity, limit the size and accuracy of current implementations. Addressing these challenges through resource-efficient designs and platform-specific optimizations is critical for advancing quantum simulators and harnessing their full potential.

In superconducting circuits, quantum computing and simulations are typically implemented using architectures based on coupled resonators and transmon qubits \cite{PhysRevA.69.062320, TransmonPaper}. Scalability is constrained by geometric limitations in the layout of circuits, even for medium-sized systems. Furthermore, the finite coherence times of transmon qubits and imperfections in gate fidelity elucidate the necessity to minimize the hardware and computational resources required for quantum simulations \cite{Kjaergaard2020}. The development of dedicated devices tailored for specific quantum simulations is an effective strategy to reduce both hardware resources and control complexity \cite{Buluta2009}.

Analog quantum simulation approaches typically utilize dedicated devices to efficiently model the target Hamiltonian \cite{Braumuller2017,Ballester2012,PhysRevX.10.021060,Flurin2017}. Particularly, previous studies have suggested implementations of the Dirac equation on circuit QED system and trapped-ion system \cite{Lamata2007, Gerritsma2010, Gerritsma2011, casanova2010, Jiang2022, Svetitsky2019, Pedernales_2013}, enabling the analog simulation of quantum relativistic phenomena such as spinor dynamics, Zitterbewegung effect and the Klein paradox \cite{Klein1929}.

In circuit QED systems, one may find two noteworthy attempts. The first maps the 1+1 dimensional Dirac Hamiltonian onto a qubit-resonator system \cite{Pedernales_2013}. Although it may appear similar in nature to the method presented in this work, that approach utilizes flux-tunable qubits in resonance with an electromagnetic mode in the resonator, which imposes stringent constraints on the system parameters. Furthermore, flux qubits have a larger footprint, and the requirement for orthogonal drives complicates fabrication constraints. Another attempt has been proposed to simulate a 3+1 dimensional Dirac spinor by coupling two qubits, thereby creating a four-level system \cite{Svetitsky2019}. Although this approach allows for the simulation of spinors in higher dimensions, it provides limited information about the simulated particle. Also, in this case, the design parameters are rigid, and thus the model Hamiltonian is restricted.

In this work, we propose a simulation method that overcomes the limitations of previous approaches in circuit QED systems \cite{Pedernales_2013, Svetitsky2019}, enabling the simulation of a wider range of scenarios in higher dimensions with fewer constraints. Our architecture trades the flux tunable qubits for Rabi dressing and realizes the qubit-resonator interaction in the dispersive regime, thus achieving the simulation with significantly alleviated parameter constraints. Through the adjustment of Hamiltonian parameters via additional control pulses, we demonstrate simulations of various quantum relativistic effects, including the one and two-dimensional Zitterbewegung effect, the Landau levels of a Dirac particle in a magnetic field, and the Klein paradox.

While similar to the Dirac Hamiltonian simulation methodology in trapped-ion systems \cite{Lamata2007, Gerritsma2010, Gerritsma2011, casanova2010, Jiang2022}, our approach has been developed to enable a wider range of simulations with fewer resources. A comparative analysis of the differences, advantages, and limitations inherent to each platform will be presented in the conclusion, based on their respective characteristics. Furthermore, we systematically identify and analyze errors arising from imperfections in the emulation of the Dirac Hamiltonian. By investigating these errors, we propose strategies to mitigate these effects \cite{Buluta2009}.

\section{\label{sec:Dirac-model}Circuit QED System for Dirac Hamiltonian Simulation: Key Components and Mapping Methods}

\begin{figure}
    \centering
    \includegraphics[width = 0.4\textwidth]{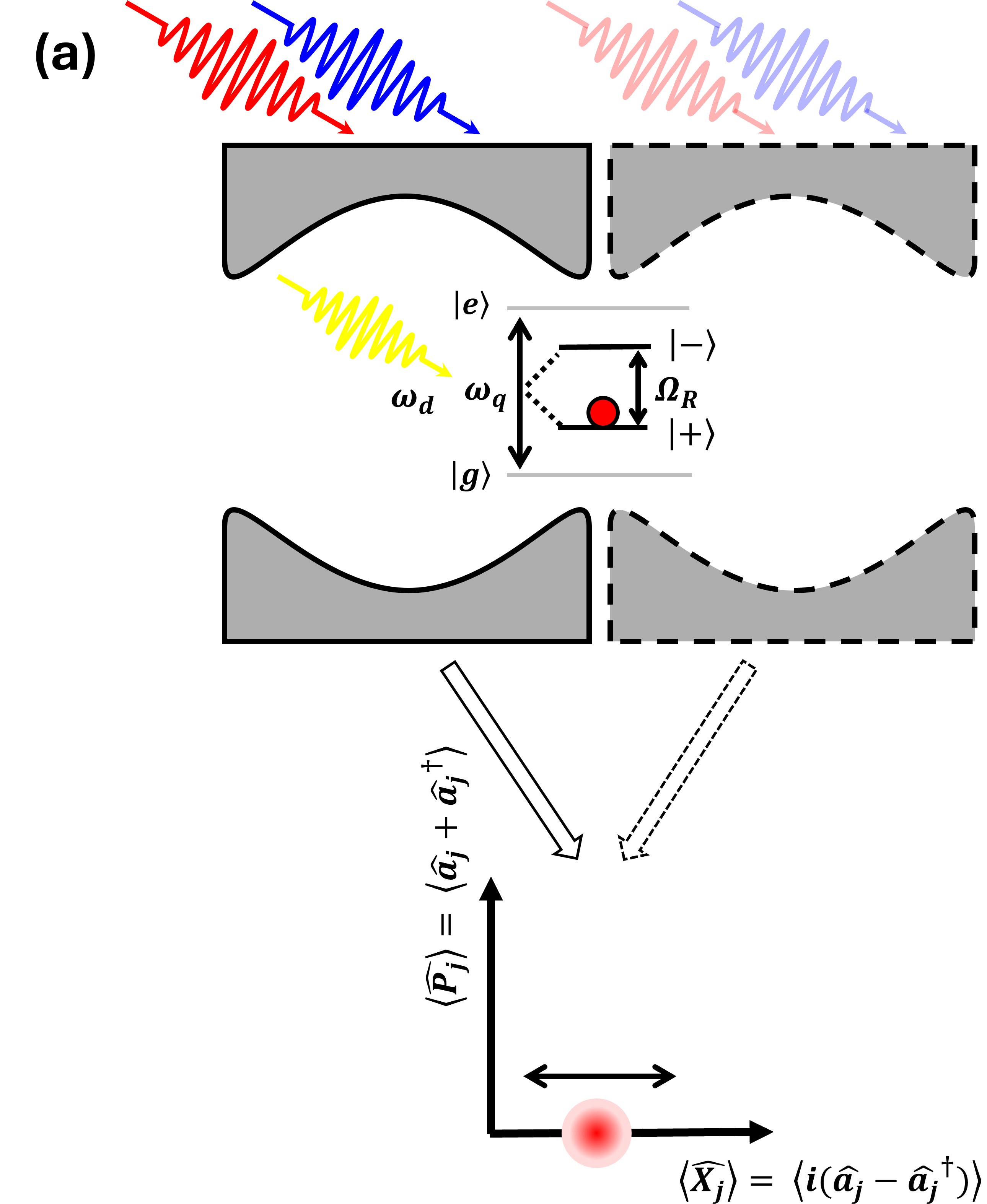}
    \includegraphics[width = 0.4\textwidth]{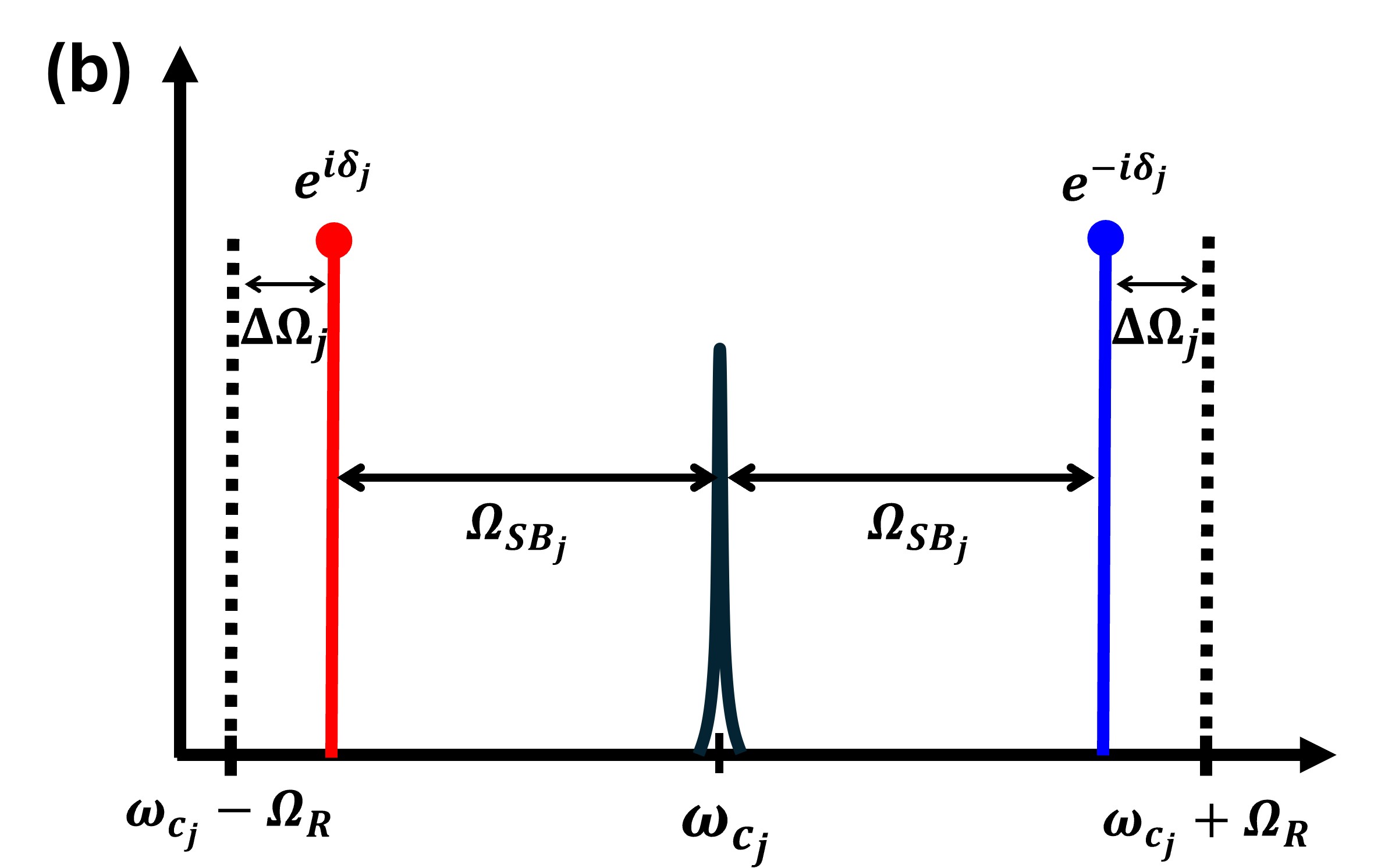}
    \caption{(a) The schematic representation of the cQED system consists of two high-Q cavities that are coupled via a superconducting qubit. Additionally, three distinct driving signals are depicted: the red and blue signals represent sideband tones for each of the high-Q cavities, while the yellow signal corresponds to a resonant Rabi drive at the qubit frequency, creating an "effective qubit." Quadratures of each cavity modes are mapped to the position and momentum operators of the spinor. (b) The configuration of the double sideband signals for the quantum simulation is shown. Red and blue sideband tones are applied at detuned frequencies from the $j_{th}$ cavity resonance frequency ($\omega_{c_j}$) by the the sideband frequency ($\Omega_{SB_j}$), with a relative phase difference of $\delta_j$.}
    \label{fig: model}
\end{figure}

The Dirac equation marks a significant milestone in physics by unifying quantum mechanics with special relativity, providing a natural description of spin-$\frac{1}{2}$ massive particles and predicting the existence of antimatter \cite{Thaller2013, Anderson1933}. For a spin-$\frac{1}{2}$ particle with rest mass $m$, the Dirac equation is expressed as \cite{Thaller2013}
\begin{equation} \label{eq: Dirac-3D} i\hbar\frac{\partial\psi}{\partial t} = (c\bm{\alpha}\cdot\bm{P} + \beta mc^2)\psi \equiv H_D\psi, \end{equation}
where $c$ is the speed of light, $\bm{P} \equiv (\hat{P_x},\hat{P_y},\hat{P_z})$ denotes the momentum operator, and 
$\bm{\alpha}$ and $\beta$ are Dirac matrices, defined using Pauli matrices and the identity operator. Generally, the wavefunction $\psi$ is the four-component Dirac spinor which includes positive and negative energy solutions, reflecting the particle-antiparticle symmetry. This paper primarily addresses the Dirac equation in 1D and 2D systems, which highlights the importance of formulating the reduced Dirac equation. Rewriting \autoref{eq: Dirac-3D} specifically for the 1D or 2D case, we obtain the following form.
\begin{equation} 
\label{eq: Dirac-1D,2D} 
i\hbar\frac{\partial\psi}{\partial t} = ((\sum_{j=1,2}^{n}{c\hat{P_j}\sigma_j}) + mc^2\sigma_z)\psi \equiv H_{nD}\psi, 
\end{equation}
where $\hat{P}_{1},\hat{P}_{2}$ indicates x, y-direction momentum operator $\hat{P}_{x},\hat{P}_y$ respectively, and $\sigma_1,\sigma_2$ indicates $\sigma_x,\sigma_y$ Pauli operator respectively. $n$ denotes the spatial dimension. In the reduced Dirac equation, which involves $2\times2$ matrix representation, $\psi$ denotes the two-component reduced Dirac spinor. This does not correspond to the two components of a spin-$\frac{1}{2}$ system; instead, it signifies a linear combination of positive and negative energy solutions of the reduced Dirac equation \cite{thaller2004,Lamata2007}.

To realize the Dirac Hamiltonian using a circuit QED system we start from the typical dispersive interaction between the cavity mode and the qubit. We then add two drives to transform the Hamiltonian into an effective Jaynes-Cummings Hamiltonian whose parameters are controlled by the phases and amplitudes of the two drives~\cite{Hacohen2016}. We then show how to map this Hamiltonian to the Dirac Hamiltonian. The two controllable drives are: a resonant Rabi drive at the qubit frequency and sideband drives detuned from the resonance frequency of the cavity mode. By applying a resonant Rabi drive to the qubit, we generate an "effective qubit," where the two-levels of the effective qubit correspond to the reduced Dirac spinor in the Dirac Hamiltonian. The dynamics of the driven qubit part of the system are described by the Hamiltonian:
\begin{equation} \label{eq: Rabi-driven qubit}
    \frac{H_q}{\hbar} = \frac{\omega_q}{2}\sigma_z + \epsilon_R(t)\sigma^{\dagger}+\epsilon_R^*(t)\sigma
\end{equation}
where $\epsilon_R(t) = \Omega_R e^{i(\omega_d t + \Delta)}$, with $\Omega_R$ as the Rabi drive amplitude and $\omega_d$ as the drive frequency. Transforming into the interaction frame with respect to $\omega_d$ by applying $U = e^{\frac{i\omega_d}{2}\sigma_z}$, and setting $\omega_d$ for resonant qubit driving, the Hamiltonian becomes $H_q' = \frac{\Omega_R}{2}\sigma_\Delta$, where $\sigma_\Delta \equiv \sigma_x\text{cos}{\Delta} + \sigma_y\text{sin}{\Delta}$. Generally, $\Delta$ is set to be 0. In this interaction frame, the new two-level system effectively acts as the reduced Dirac spinor, as illustrated in Fig. 1(a).

Additionally, a double sideband driving is applied at frequencies detuned by $\pm\Omega_{SB_j}$ from the cavity resonance frequency $\omega_{c_j}$ with identical amplitude $\alpha_j$ and relative phases $\delta_j$. The frequency-domain representation of this driving scheme is depicted in Fig. 1(b). The double sideband-driven cavity Hamiltonian can be expressed as:
\begin{equation}
    \frac{H_{c}}{\hbar} = \sum_{j=1}^n~[\omega_{c_j}\hat{a_j}^\dagger \hat{a_j} + \varepsilon_j(t)\hat{a_j}^\dagger+\varepsilon_j(t)^*\hat{a_j}]
\end{equation}
where $\hat{a_j}$, $\hat{a_j}^{\dagger}$ are ladder operators of the $j_{th}$ cavity mode, and
\begin{equation}
\begin{aligned}
\label{eq: Double_sidebands}
    \varepsilon_j(t) 
    &= \frac{\alpha_j\Omega_{SB_j}}{2}(e^{-i(\omega_{c_j}+\Omega_{SB_j})t-\delta_j}-e^{-i(\omega_{c_j}-\Omega_{SB_j})t+\delta_j}).
    \end{aligned}
\end{equation}
When $\Omega_{SB_j} = \Omega_{R}$, the resulting Raman scattering produces photons at the cavity’s resonance frequency, facilitating processes such as cavity-mediated cooling or heating of the effective qubit \cite{PhysRevLett.109.183602, Hacohen2016}. On the other hand, when $\Omega_{R}-\Omega_{SB_j} \equiv \Delta\Omega_j \neq 0$, an additional z-axis rotation is introduced to the effective qubit. These dynamics can be expressed in a displaced frame as:
\begin{equation} 
\label{eq: Dirac_cqed}
    \frac{H^{'}_{nD}}{\hbar} = \sum_{j=1}^{n}~[\frac{{\chi_j}{\alpha_j}}{4}(\hat{a_j}^\dagger+\hat{a_j})\sigma_{\delta_j} + \frac{{\Delta}{\Omega_j}}{2}\sigma_z]
\end{equation}
where $\chi_j$ is the coupling strength between a qubit and $j_{th}$ cavity mode, $\sigma_{\delta_j} = \sigma e^{i\delta_j} + \sigma^{\dagger}e^{-i\delta_j}$, and $\sigma$, $\sigma^{\dagger}$ are lowering, raising operators of effective qubit. The detailed derivation of \autoref{eq: Dirac_cqed} is provided in Appendix A.

This circuit QED Dirac Hamiltonian \autoref{eq: Dirac_cqed} is equivalent to the reduced Dirac Hamiltonian \autoref{eq: Dirac-1D,2D} where the parameters can be controlled by the drives. First, the speed of light $c$ is replaced by $\frac{\chi_j\alpha_j}{4}$, and rest mass $mc^2$ by $\sum_{j=1}^{n}{\frac{\Delta\Omega_j}{2}}$. Second, the Dirac spinor's momentum operator $\hat{P_{x}}$ and position operator $\hat{X}$ in the $x$-direction are mapped onto the quadratures of the 1st cavity mode $\hat{a}_{1}^\dagger+\hat{a}_{1}$ and $i(\hat{a}_{1}-\hat{a}_{1}^\dagger)$ respectively. Similarly, the $y$-direction operators can be mapped with those of the 2nd cavity mode. Finally, Pauli operators of the Dirac spinor $\sigma_x, \sigma_y$ corresponds to Pauli operators of the effective qubit $\sigma_{\delta_1}, \sigma_{\delta_2}$. In general, the relative phases are set to be $\delta_1 =0$, $\delta_2=\pi/2$.

By modifying the double sideband driving term in \autoref{eq: Double_sidebands}, the Dirac equation can be extended to include the effects of magnetic or electrostatic potentials. Furthermore, incorporating an additional cavity mode enables simulations in two dimensions. Since the parameters of the circuit QED system directly correspond to those of the Dirac Hamiltonian, simulations under a wide range of conditions can be performed by simply adjusting the simulation parameters.

The procedure for simulating specific quantum relativistic effects via the Dirac Hamiltonian in various environments is as follows: The initial states of the effective qubit and the cavity mode are prepared. Subsequently, the desired Dirac Hamiltonian is implemented using the described methodologies, allowing the system to undergo time evolution for a specified duration. The resulting quantum state is an entangled state of the cavity mode and the effective qubit, from which relevant outcomes can be extracted via measurement.

The measurement protocol for the resulting state would involve the following steps: (a) Measure the effective qubit, causing its state to collapse and disentangle from the cavity modes. (b) then a characteristic function measurement of the relevant cavity modes can be performed \cite{PhysRevX.14.011055}. (c) By repeating steps (a) and (b), a classical mixture containing all information, except the relative phase between the cavity mode state and the qubit can be obtained. This protocol enables determination of the position and momentum of the Dirac spinor, enabling a comprehensive analysis of the simulated quantum relativistic effects.

In the following section, we present methods for simulating three distinct scenarios: the motion of a Dirac particle in free space, its behavior in a magnetic field, and the Klein paradox in the presence of an electric field.

\section{Numerical characterization of the system}

\subsection{\label{sec:free-Dirac}Free Dirac particle}

\begin{figure}
    \centering
    \includegraphics[width = 0.4\textwidth]{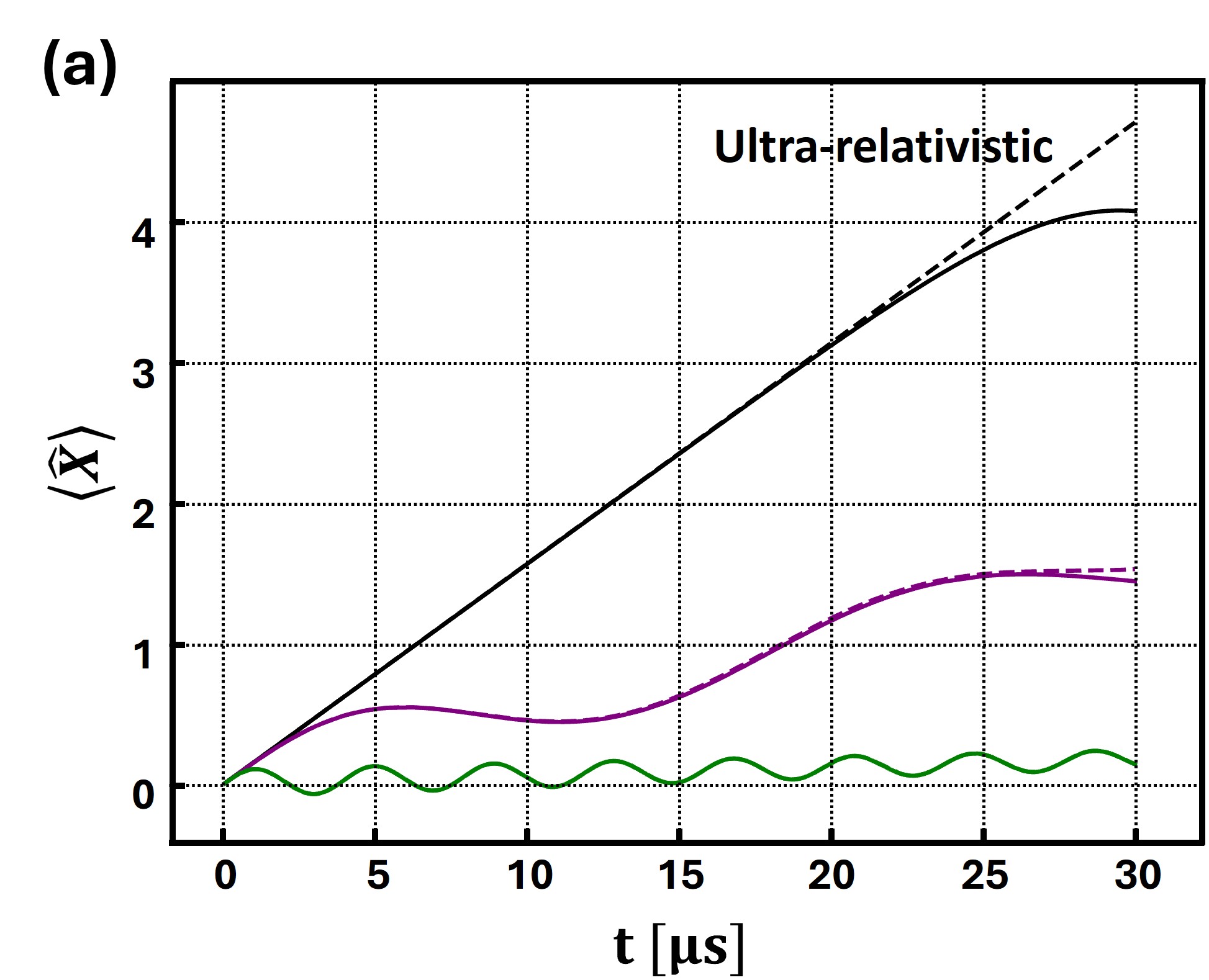}
    \includegraphics[width=0.4\textwidth]{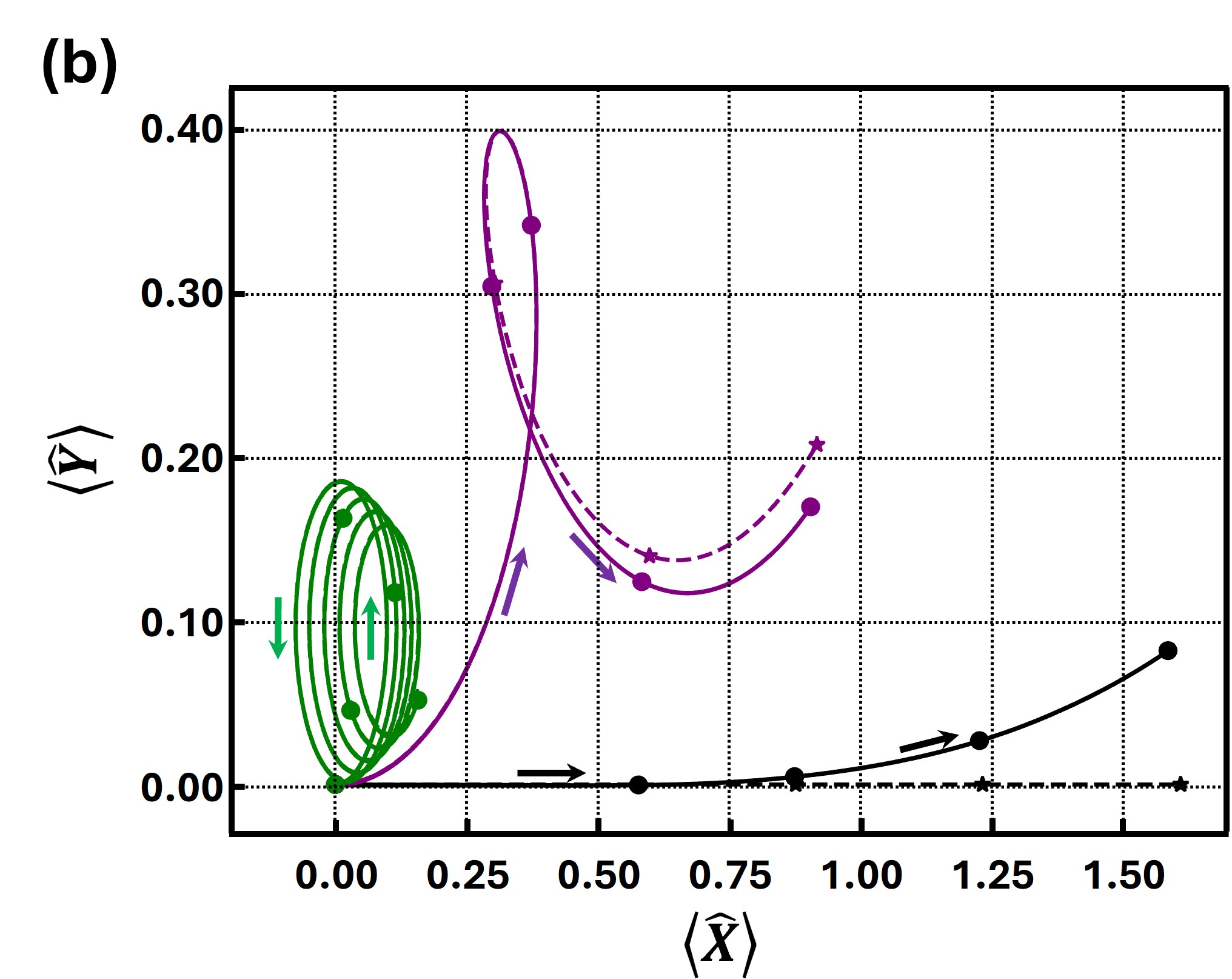}
    \includegraphics[width=0.4\textwidth]{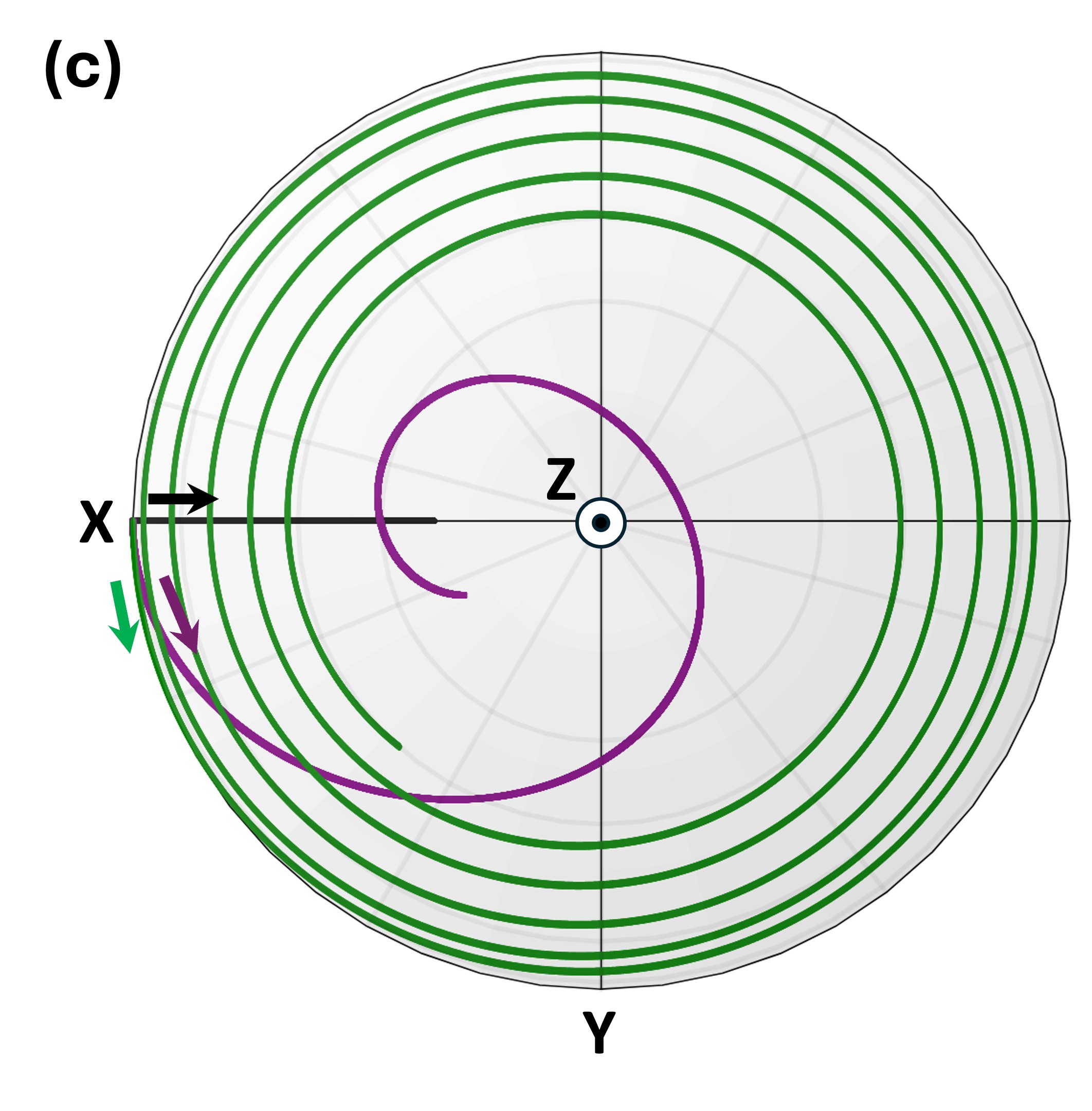}
    \caption{(a) and (b) depict the motion of a free Dirac particle in 1D and 2D, respectively, as a function of particle mass ($\Delta\Omega$). The mass is represented by different colors in (a), (b), and (c): Black, Purple, and Green correspond to $\frac{\Delta\Omega}{2\pi}$ values of 0 MHz, 0.05 MHz, and 0.25 MHz, respectively. The dashed lines represent the ideal Dirac Hamiltonian simulation, while the bold lines represent the full Hamiltonian simulation. (b) illustrates the 2D trajectory of the particle starting from (0,0) at $T=0\mu$s, with each point on the trajectory separated by a 5$\mu$s interval, covering a total time of 20$\mu$s. (c) illustrates the trajectory of the effective qubit state, initialized in $\ket{+}$, during the evolution of a 2D Dirac particle over a 20 $\mu s$ period.}
    \label{fig: zitterbewegung}
\end{figure}

In the case of a free particle, Zitterbewegung, the characteristic jittery motion of a massive particle, is expected. These oscillations are understood to result from the interference between the positive and negative energy components of the spinor \cite{Gerritsma2010}. The general Hamiltonian for a free Dirac particle is described by \autoref{eq: Dirac-3D}. The Hamiltonian, when reduced to one or two dimensions, results in \autoref{eq: Dirac-1D,2D}. As discussed in the previous section, by applying Rabi driving to the qubit and symmetric double sideband driving to the cavity, we can derive the Hamiltonian in the form of \autoref{eq: Dirac_cqed}, which corresponds to \autoref{eq: Dirac-1D,2D}.

Fig. 2(a) illustrates the motion of a free Dirac particle in 1D, with the particle's mass represented by different colors. In all cases, the initial state of the cavity is the vacuum state, while the initial state of the effective qubit is a superposition of the ground and excited states, denoted as the $\ket{+}$ state. ($\ket{0}_{c_1}\otimes\ket{+}_q$) In the case of a 1D massless particle $\Delta\Omega = 0$ (black lines), since $[\sigma_x, H^{'}_{1D}] = \frac{d\sigma_x}{dt} = 0$ and the initial state of the effective qubit is an eigenstate of $\sigma_x$, the qubit state remains unchanged over time. Furthermore, from $[\hat{X}, H^{'}_{1D}] = \frac{d\hat{X}}{dt} = \frac{\chi_1\alpha_1}{4} =c$, we can determine the velocity of the massless particle. In a similar manner, for the case of a 1D massive particle ($\Delta \Omega \neq 0$), as shown by the purple and green lines, the commutation relation $[\sigma_x, H^{'}_{1D}] = \frac{d\sigma_x}{dt} = -\Omega\sigma_y$ indicates that the effective qubit state undergoes rotation around the z-axis. As a result, the cavity's coherent state will oscillate in the X quadrature direction, depending on the state of the effective qubit. This manifests as jittering motion.

The commutation relation $[\hat{P_x}, H^{'}_{1D}] = \frac{d\hat{P_x}}{dt} = 0$
indicates that the momentum of the particle is conserved throughout its motion. Moreover, the assumption that the cavity's initial state is the vacuum state implies that both the initial position and momentum of the Dirac spinor are zero. As a result, while the initial momentum is zero and remains conserved during time evolution, the particle’s position evolves in a manner that may seem counterintuitive. This phenomenon arises from the fact that in the quantum relativistic regime, the system follows the $H^{'}_{1D} \sim \hat{P_x}$. Consequently, as the particle's mass increases and the system transitions towards the non-relativistic regime, the particle's motion converges to a rest position at zero. Thus, the Zitterbewegung motion is most clearly observed in the intermediate relativistic regime, which lies between the ultra-relativistic and non-relativistic limits.

Fig. 2(b) presents the time evolution of a free Dirac particle in two dimensions. The two cavity modes are initialized in the vacuum state, while the effective qubit is prepared in the $\ket{+}$ state ($\ket{0}_{c_1}\otimes\ket{0}_{c_2}\otimes\ket{+}_q$). Consequently, all simulations of mass-dependent motion start from the origin, with an initial average momentum of zero. Similarly with the 1D case, where $\frac{d\hat{P_x}}{dt} = [\hat{P_x}, H^{'}_{2D}] \sim [\hat{P_x}, \hat{P_y}] = 0$ (and similarly, $\frac{d\hat{P_y}}{dt} = 0$), average momentum is constant during the motion. In the intermediate relativistic regime, the Zitterbewegung effect manifests as spiral motion in two dimensions. When the 2D spiral motion is projected onto a specific axis, the resulting behavior resembles the Zitterbewegung observed in 1D systems.

Another important observation is that the amplitude of the spiral motion decreases over time. As time progresses, the effective qubit state and the cavity state become increasingly entangled due to effective qubit state dependent cavity state displacement. Consequently, at each time step the state appears to rotate along an axis perpendicular to the XY plane of the Bloch sphere, with a gradual reduction in purity, as shown in Fig. 2(c). This reduction in purity results in a narrowing of the possible ranges for $\langle \sigma_x \rangle$ and $\langle \sigma_y \rangle$, and subsequently leads to a decrease in the amplitude of the cavity state displacement. The slower propagation speed of a massless particle in 2D compared to 1D can be explained by the decreasing purity over time.

All simulations are presented by comparing the results obtained from the ideal Dirac Hamiltonian and those from the full Hamiltonian in the circuit QED system. As the dynamics evolve over time, we observe increasing deviations in the simulation results. This arises because, over time, photons accumulate in the cavity, causing the Rotating Wave Approximation (RWA) to no longer hold. As a consequence, additional Hamiltonian terms emerge. These terms can be computed using the Magnus expansion \cite{Magnus1954}, yielding a second-order correction term given by: $H^{(2)}_{\text{Magnus}} \approx \frac{\chi_j^2}{4\Omega_{SB_j}}(\hat{a_j}^\dagger \hat{a_j})^2\sigma_z$.
To minimize the impact of these unwanted additional terms, it is necessary to increase $\Omega_{SB}$, however in experiments this parameter is limited to 10s of MHz. A detailed derivation of these calculations is provided in Appendix C.

\subsection{\label{sec:magnetic-Dirac}Dirac particle in a magnetic field}

\begin{figure}
    \centering
    \includegraphics[width = 0.45\textwidth]{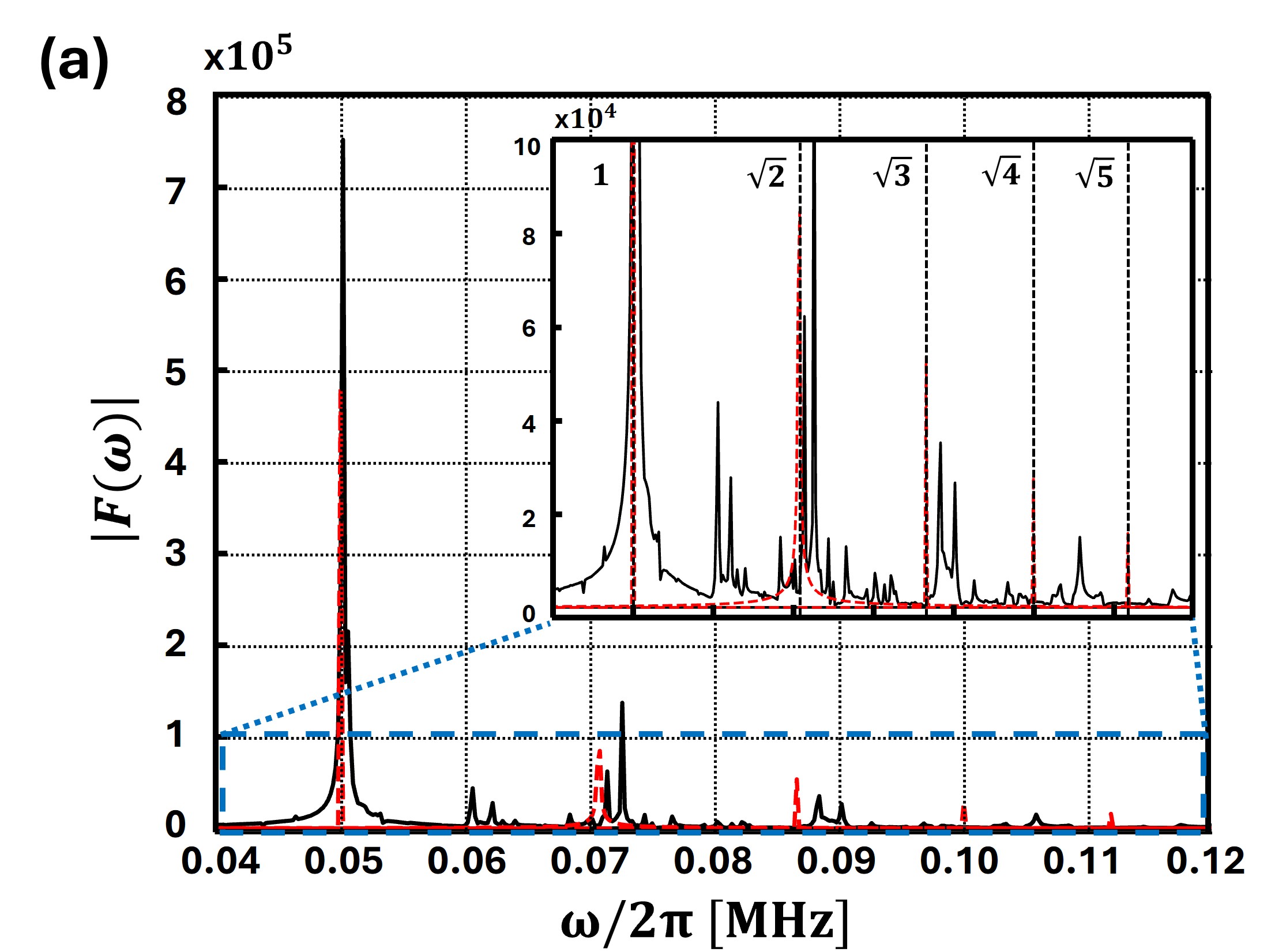}
    \includegraphics[width = 0.43\textwidth]{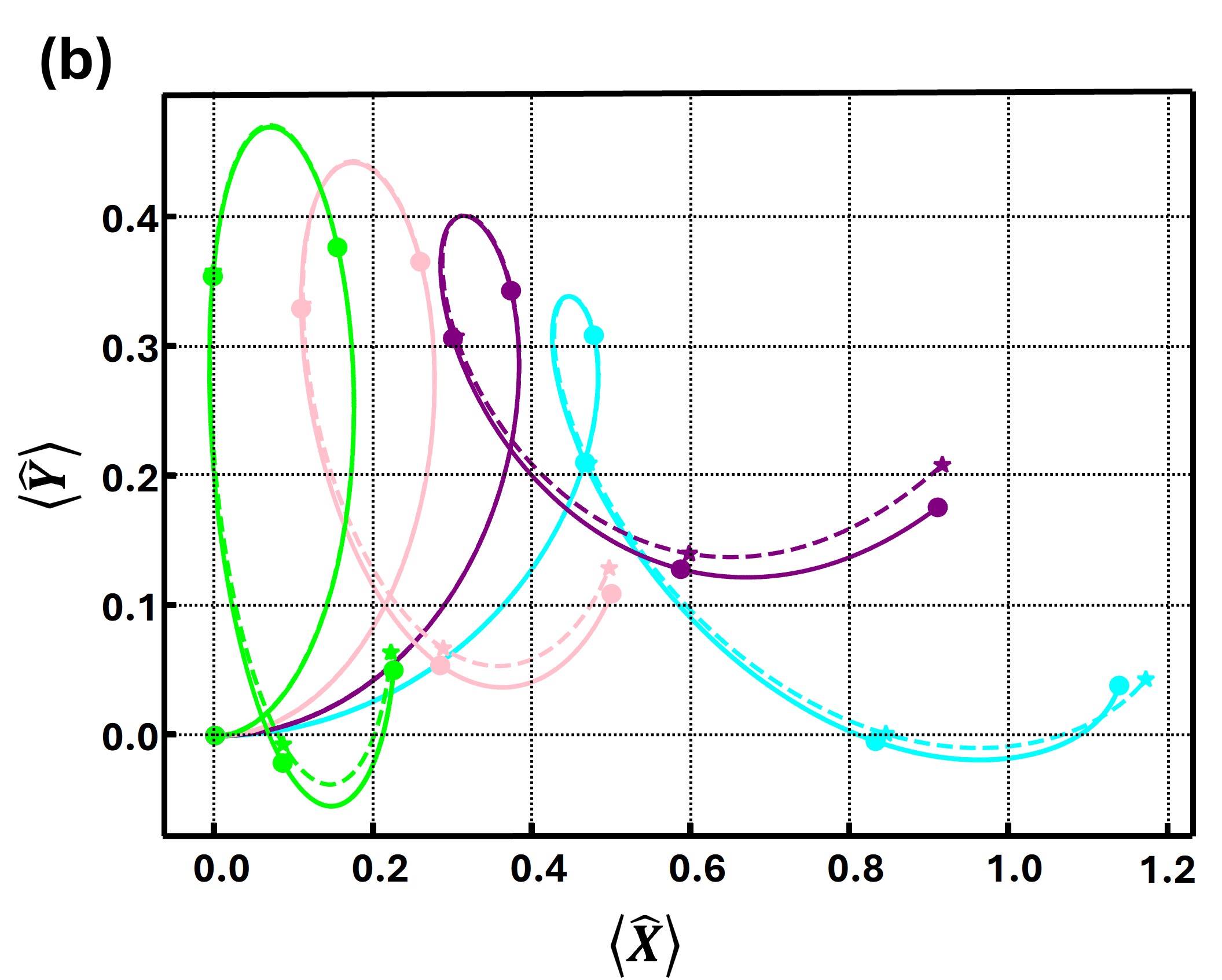}
    \caption{(a) The Fourier transform result of the qubit expectation value $\langle\sigma_z(t)\rangle$ reveals the energy spectrum of a massless Dirac particle in a magnetic field. The relativistic Landau energy levels are given by the relation $E_n \propto \sqrt{n}$, as shown in the inset. The red dashed line represents the result of ideal Dirac Hamiltonian simulation, black bold line represents the result of full system Hamiltonian simulation. (b) The time trajectory of the Dirac particle with same mass ($\frac{\Delta\Omega}{2\pi} = 0.05\text{MHz}$) in a perpendicular magnetic field is illustrated, with varying colors representing different strengths and directions of the magnetic field: cyan ($\Delta\alpha_j$ = 0.5), purple ($\Delta\alpha_j$ = 0), pink ($\Delta\alpha_j$ = -0.5), and lime ($\Delta\alpha_j$ = -1), while $\alpha_j$ is fixed at 1. Each trajectory begins at (0,0) at $T=0\mu$s, with a total evolution time of 20 µs and a time interval of 5 µs between each point. Each dashed and solid line represents the same meaning as in Fig. 2.}
    \label{fig: magnetic}
\end{figure}

The Hamiltonian for a 3D charged Dirac particle in a magnetic field is expressed as $i\hbar\frac{\partial\psi}{\partial t} = (c\bm{\alpha}\cdot(\bm{p}-e\bm{A}) + \beta mc^2)\psi$ from \autoref{eq: Dirac-3D} \cite{casanova2010}. In the case of a magnetic field applied perpendicular to a 2D plane, the Dirac Hamiltonian becomes 
\begin{equation} \label{eq: Dirac in magnetic} 
H_{2D,\text{mag}} = c\sigma_x \hat{P_x} + c\sigma_y(\hat{P_y} - eB\hat{X}) + mc^2\sigma_z 
\end{equation} 
where $\bm{p} = (\hat{P_x},\hat{P_y},0)$, $\bm{A} = (0, B\hat{X}, 0)$. To realize this Hamiltonian in a circuit QED system, an asymmetric double sideband drive, denoted as $\varepsilon'_j$, must be applied to one of the two modes, while a symmetric double sideband drive, as given by \autoref{eq: Double_sidebands}, is required for the other mode, where $\varepsilon'_j = \Omega_{SB_j}(-\frac{\Delta\alpha_j}{2}e^{i(\Omega_{SB_j}t+\delta_j)}-i\alpha_j\text{sin}(\Omega_{SB_j}t+\delta_j))e^{-i\omega_{c_j}t}$. 
Through the Hamiltonian mapping, the magnetic field $eB$ is related to $\frac{2\Delta\alpha_j}{2\alpha_j+\Delta\alpha_j}$. Therefore, the magnetic field can be controlled by adjusting $\Delta\alpha_j$.

Landau energy levels describe the quantization of the cyclotron motion of charged particles in a perpendicular magnetic field. In the non-relativistic regime ($m \gg eB$), the discrete energy levels are given by $E_n = \frac{neB}{m}$ \cite{Landau1930}, whereas in the relativistic regime, they follow $E_n = \sqrt{m^2+2neB}$ \cite{Lamata2011}. From the Fourier transform of the time-dependent qubit signal $\langle\sigma_z(t)\rangle$, shown in Fig. 3(a), the Landau energy levels for a Weyl (massless) particle are determined \cite{Jiang2022}. To simulate the Hamiltonian, an initial state is prepared as $\ket{0}_{c_1}\otimes\ket{0}_{c_2}\otimes\ket{+}_q$. Using the simulation parameters, such as $\frac{\chi_j}{2\pi} = 0.1\text{MHz}$ and $\Delta\alpha_j = -1$, the energy difference between the positive and negative Weyl spinor states can be calculated as $\Delta E_n = 2\sqrt{2neB} = \frac{\chi_j\Delta\alpha_j}{2}\sqrt{n}$. As shown in Fig. 3(a), the first quantized energy level is located at 0.05 MHz. Additional energy peaks, spaced by factors of $\sqrt{2}$, $\sqrt{3}$, $\sqrt{4}$, and $\sqrt{5}$ from the first level, are consistent with the ideal Dirac Hamiltonian simulation, confirming the theoretical relation $E_n \propto \sqrt{n}$. While the results of full Dirac Hamiltonian simulation in circuit QED system exhibit similar trends, deviations from the ideal case arise due to the limitations of the rotating wave approximation (RWA). For a Dirac particle with mass, the frequency difference between the Landau energy levels will increase by the amount corresponding to the added mass (Detailed in Appendix D).

Fig. 3(b) depicts the motion of a 2D Dirac particle in the presence of a magnetic field. The purple trajectory corresponds to the motion of a free Dirac particle without a magnetic field, identical to the one shown in Fig. 2(b). Using this trajectory as a reference, it can be observed that charged particles experience Lorentz forces due to the magnetic field's direction and magnitude, causing their motion to follow a spiral path, consistent with classical mechanics. Moreover, over time, discrepancies between the two simulations become noticeable, arising from the limitations of the rotating wave approximation (RWA).

\subsection{\label{sec:Klein}Klein paradox}

\begin{figure*}
    \centering
    \includegraphics[width = 0.9\textwidth]{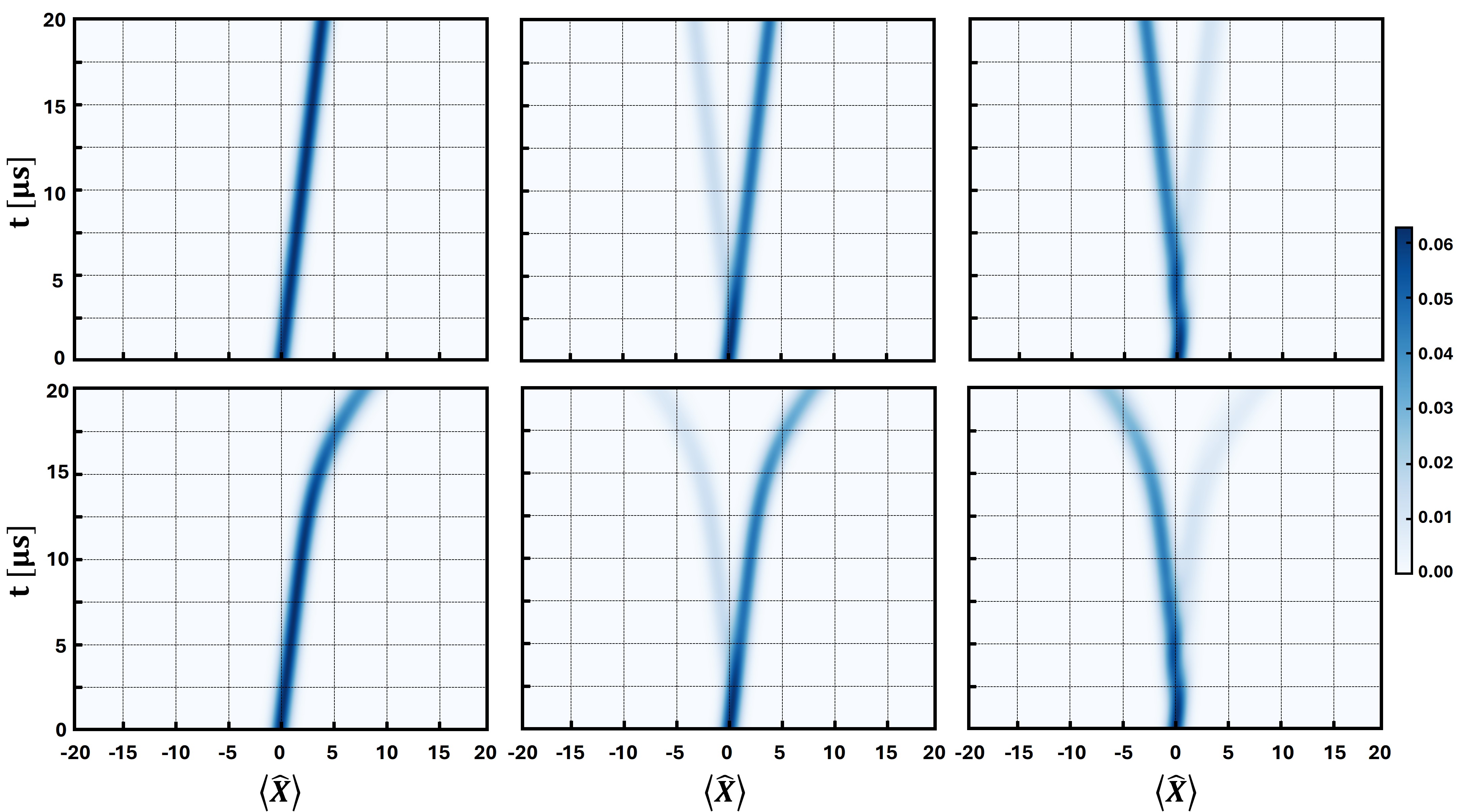}
    \caption{The scattering behaviors of three distinct masses  ($\frac{\Delta\Omega}{2\pi} = 0, 0.05, 0.15 \text{MHz}$, from left to right) are investigated in the presence of a linear electrostatic potential $g\hat{X}$, where $\frac{g}{2\pi} = 0.1\text{MHz}$.  The total probability density is expressed as a function of both space $\langle\hat{X}\rangle$ and time $t$. The upper row of graphs corresponds to trajectories obtained from simulations using the Ideal Dirac Hamiltonian, while the lower row presents trajectories derived from the Dirac Hamiltonian in the context of a circuit QED system.}
    \label{fig: Klein paradox-1}
\end{figure*}

The 1D Dirac Hamiltonian in a linear electrostatic potential is expressed as
\begin{equation} \label{eq: Dirac-elec}
    H_{1D,\text{elec}} = c\sigma_x \hat{P_x} + mc^2\sigma_z + g\hat{X}.
\end{equation}
To realize this Hamiltonian in a circuit QED system, a cavity mode resonant drive needs to be applied additionally, alongside the symmetric sideband drive, as described in \autoref{eq: Double_sidebands}. The detailed derivation is provided in Appendix A. According to the Hamiltonian mapping, the slope $g$ of the electrostatic potential can be controlled by adjusting the magnitude of the cavity mode resonant driving. 

An intriguing phenomenon that arises when a Dirac particle encounters a potential wall is the Klein paradox. This effect occurs when the potential wall is significantly higher than the kinetic energy of the Dirac particle, allowing the particle to propagate beyond the wall. Specifically, propagation becomes possible when the potential height $V$ exceeds the particle's kinetic energy $E-mc^2$ by at least $2mc^2$ \cite{Gerritsma2011}. In the context of quantum field theory, this phenomenon is often interpreted as electron-positron pair production resulting from the collision of an electron with the potential wall \cite{PhysRev.82.664, Hund1941}.

The solutions for the negative and positive energy states of a Dirac spinor are given by $E_{\pm}=\pm\sqrt{c^2p^2+m^2c^4}$, which represents a hyperbolic curve in momentum-energy space, centered at the origin. In the presence of an external linear potential, the momentum becomes time-dependent, such that $\frac{dp}{dt}\neq 0$, and the Klein paradox can be partially explained through Landau-Zener tunneling \cite{PhysRevA.23.3107}. For ultra-relativistic particles, where $E \sim \pm pc$, the phenomenon can be entirely attributed to Landau-Zener tunneling, with the tunneling probability given by $P_{tunnel} = e^{-2\pi\Gamma}$, where $\Gamma = \frac{m^2c^3}{2\hbar g}$ \cite{Gerritsma2011}. In general cases, the tunneling probability is determined through numerical calculations.

Fig. 4 presents simulation results for the transmission of a Dirac particle through a potential wall for varying particle mass. As the mass increases, the gap between the two energy branches grows, causing the particle to adiabatically follow the positive energy branch. This leads to an increase in the reflection rate. This behavior can also be observed by adjusting the slope of the linear electrostatic potential. We simulated the trajectory of a Dirac particle with initial momentum $\langle\hat{P_x}\rangle = \alpha_1 (=1)$ and initial position $\langle\hat{X}\rangle = 0$ in an environment with a linear electrostatic potential $g\hat{X}$, where $\frac{g}{2\pi} = 0.1\text{MHz}$ over a 20$\mu s$ period. Additional details, including explanations of the physical scenario, can be found in the Appendix E. The three graphs in the upper row correspond to simulations of the ideal Dirac Hamiltonian, while the three graphs in the lower row show simulation expected from a circuit QED system. As illustrated in the graphs, after collisions with the potential wall, spinors with positive and negative energy exhibit opposite directional motion. The transmitted probabilities for the three cases are 1, 0.75, and 0.23, respectively. Significant discrepancies between the two methods emerge after around 12.5$\mu s$. This can be attributed to the limitations of applying the rotating wave approximation (RWA) due to photon accumulation within the cavity. Despite the significant divergence in the trajectories after 12.5$\mu s$, the transmitted probabilities remain unaffected.

\section{Conclusions}

In this paper, we have presented a method for simulating a range of quantum relativistic effects using a dedicated device that requires minimal resources. The suggested method is designed to perform these simulations with a single cavity featuring multiple modes and a single qubit that can couple to each cavity mode \cite{PhysRevX.14.011055,chakram2022multimode}. The essence of the method uses Rabi driving to generate an effective qubit, and applied sideband driving to the cavity mode to gain full control of the interactions. 

Our approach shares a few similarities with the methodology proposed in trapped-ion systems \cite{Lamata2007, Gerritsma2010, Gerritsma2011, casanova2010, Jiang2022}. Specifically, we map the Dirac spinor to the states of a Rabi-driven effective qubit and the position and momentum operators of the Dirac Hamiltonian to the quadratures of the cavity mode. Likewise, in trapped-ion systems, the Dirac spinor is encoded in the internal two-level states of an ion, while the position and momentum operators are represented by the quadratures of the ion’s vibrational mode \cite{Lamata2007}. In essence, the Rabi dressing allows us to emulate the dynamics in the ion-trap with the different energy scales in circuit-QED (see \cite{Lu2022} for a detailed explanation).

Despite these similarities, the two platforms exhibit notable differences in two key aspects. First, the energy scales govern the timescale for the simulation. In trapped-ion systems, the coupling rate is on the order of tens of kHz \cite{Gerritsma2010, Gerritsma2011, Jiang2022}, whereas our approach achieves coupling rates in the range of hundreds of kHz to a few MHz. As a result, the simulation time differs substantially, with trapped-ion implementations typically requiring hundreds of microseconds, while our method operates on the order of tens of microseconds, as reflected in the respective results. The second major difference lies in the resources required for Hamiltonian construction \cite{casanova2010}. In trapped-ion systems, certain conditions must be met to realize a Dirac Hamiltonian with a specific potential. A massless (Weyl) Dirac particle Hamiltonian in a particular potential can be achieved by applying two additional light fields to a single ion \cite{Jiang2022}, and simulating a massive Dirac particle requires an auxiliary ion with blue and red sidebands to generate the desired potential \cite{Gerritsma2011}. In contrast, our approach allows for the simulation of all cases using only a single qubit, reducing the required hardware resources.


This approach enables relatively fast simulations and supports a broader range of scenarios with minimal hardware requirements. However, a key limitation arises as the photon number in the system increases, causing the Rotating Wave Approximation (RWA) to gradually break down. Consequently, in all simulations of quantum-relativistic effects presented here, we observe error accumulation over time (see Appendix C). This limitation can be mitigated by carefully tuning key parameters. The Hamiltonian term responsible for the validity of the RWA is given by $\frac{\chi^2n^2}{4\Omega_{SB}}$ (see Appendix C). To suppress this term, three main strategies can be employed: increasing $\Omega_{SB}$, reducing the coupling strength $\chi$, and lowering the photon number within the cavity.

First, increasing $\Omega_{SB}$ helps mitigate RWA errors while preserving the simulation, provided that $\Omega_R$ is adjusted accordingly to maintain $\Delta\Omega$. However, experimental constraints on qubit drive power limit how much $\Omega_R$ can be increased. Additionally, reducing $\chi$ and the cavity photon number effectively suppresses RWA errors due to their quadratic dependence, but at the cost of decreasing the coefficient in \autoref{eq: Dirac_cqed}, leading to a linear increase in simulation time. Despite this trade-off, the quadratic suppression of RWA errors makes this approach promising.

Another challenge is the finite coherence time in real systems, which can limit the fidelity of quantum simulations. However, since the typical quantum simulation time is shorter than the coherence time, this issue can largely be mitigated. Further improvements are possible by enhancing the quality of transmons and cavities. Notably, state-of-the-art technology has demonstrated cavity mode lifetimes on the order of tens of milliseconds \cite{milul2023superconducting}, whereas qubit lifetimes remain more limited, approaching a millisecond \cite{tuokkola2024}. While the simulations presented in this work use fixed parameter values, optimizing these parameters remains crucial for maximizing overall performance.


Achieving reliable and consistent simulation results with a dedicated device requiring minimal resources in a circuit QED system is a promising outcome. Furthermore, by incorporating additional cavity modes and qubits, it is expected that simulations of the Dirac Hamiltonian in three-dimensional or other complex potential environments will become feasible, as suggested by the theoretical studies on trapped ion systems \cite{Lamata2007, casanova2010}. Dirac particle quantum simulator could serve as an alternative to high-energy relativistic particle physics experiments, eliminating the need for extreme energy scales. Additionally, as electrons in two-dimensional materials such as graphene exhibit massless Dirac-like behavior, this simulation approach could provide valuable insights into graphene’s electronic properties, including the anomalous quantum Hall effect \cite{Novoselov2005}. Beyond the implementation of the Dirac Hamiltonian, it is anticipated that other physical models can be simulated by suitably modifying the dedicated device. Given that the system supports the coupling of multiple cavity modes with multiple qubits, it holds potential for simulating high-dimensional quantum walks \cite{Flurin2017}, where each mode corresponds to a specific direction of particle motion, as well as the Dicke model \cite{Lamata2017, Mezzacapo2014}, where multiple qubits with identical frequencies interact with a single cavity mode.

\begin{acknowledgments}
This research was supported by the Israeli Science Foundation (ISF), Pazi foundation, and Technion's Helen Diller Quantum Center. This research was supported by Creation of the quantum information science R\&D ecosystem(based on human resources)(2022M3H3A109893113) through the National Research Foundation of Korea(NRF) funded by the Korean government (Ministry of Science and ICT(MSIT))

\setlength{\parskip}{20pt}

* These authors contributed equally to this work.
\end{acknowledgments}

\appendix \label{Appendix}

\section{Details of the Dirac Hamiltonian mapping} \label{Appendix A}

\subsection{System's Hamiltonian and the general mapping scheme}
The system comprises a Rabi driven qubit interacting dispersively with a driven multi-mode cavity. The Hamiltonian is therefore,
\begin{equation} \label{eq: System_H}
\begin{aligned}
    \frac{H}{\hbar} &= \frac{H_q + H_c + H_I}{\hbar}\\
    &\frac{H_{q}}{\hbar} = \frac{\omega_q}{2}\sigma_z + \frac{\Omega_R}{2}\sigma_x\text{cos}(\omega_dt)\\
    &\frac{H_{c}}{\hbar} = \sum_{j=\{1,2\}}(\omega_{c_j}\hat{a_j}^\dagger \hat{a_j} + \epsilon_j(t)\hat{a_j}^\dagger+\epsilon_j(t)^*\hat{a_j})\\
    &\frac{H_{I}}{\hbar} = \sum_{j=\{1,2\}}\frac{\chi_j}{2}\sigma_z \hat{a_j}^\dagger \hat{a_j}.
    \end{aligned}
\end{equation}
Where $H_q$ describes the Rabi driven qubit, $H_c$ the driven cavity modes, and $H_I$ is the dispersive interaction term. The cavity drive $\epsilon_j(t)$ consists of a sum of sideband modulations around the cavity's $j$-th resonance, 
\begin{equation}
\begin{aligned}
    \epsilon_j(t)=
    \sum_n&\sqrt{\Omega^2_{j,n}+\frac{\kappa^2_j}{4}}\frac{\alpha_{j,n}}{2}\\
    &e^{-i(\Omega_{j,n}t+\delta_{j,n}-arctan(\frac{\kappa_j}{2\Omega_{j,n}}))}e^{-i\omega_{c_j}t},
    \end{aligned}
\end{equation}
where $n$ is the index of sideband, the amplitudes $\alpha_{j,n}$, frequencies $\Omega_{j,n}$ and phases $\delta_{j,n}$ determine the shape of the emerging Dirac Hamiltonian. $\kappa_j$ is the damping rate of the cavity at the respective resonance. This is the general form of \autoref{eq: Double_sidebands} in the main text. In the following derivations, we assume that $\kappa_j$ is much smaller than $\Omega_{j,n}$, and thus $\kappa_j$ will be neglected.

The Hamiltonian $H$ in the Circuit QED system is unitary-transformed into the form of the Dirac Hamiltonian through the following process:
\begin{enumerate}
    \item We transform into the interaction picture w.r.t the $j_{th}$ cavity resonance frequency by applying the unitary $U_1=e^{i\omega_{c_j} \hat{a_j}^\dagger \hat{a_j} t/2}$.
    \item We transform into the driven qubit frame by applying the unitary $U_2=e^{i\omega_d\sigma_z t/2}$.
    \item We transform into the displaced cavity frame by applying the unitary $U_3=e^{\alpha_j(t)\hat{a_j}^\dagger-\alpha_j^*(t)\hat{a_j}}$, where $\alpha_j(t)$ is the solution to the classical driven oscillator equation.
    \item We set $\omega_d$ to cancel out all the constants multiplied by $\sigma_z$ (i.e. choosing resonant Rabi drive).
    \item We transform into the Hadamard frame ($\sigma_z\leftrightarrow\sigma_x$) and then to the sidebands frame by applying the unitary $U_4=e^{i\Omega_{SB_j} \sigma_z t/2}$.
    \item We apply rotating wave approximation(RWA).
\end{enumerate}

\subsection{Free Dirac particle} \label{app: free}
To simulate the free Dirac particle we apply a symmetric pair of sidebands which is well-described in Figure 1(b), i.e. 

\begin{equation}
\begin{aligned}
    \epsilon_j(t) &= \frac{\alpha_j}{2}\Omega_{SB_j}(e^{-i(\Omega_{SB_j}t+\delta_j)}
    -e^{i(\Omega_{SB_j}t+\delta_j)})e^{-i\omega_{c_j}t}\\
    &= -i\alpha_j\Omega_{SB_j}\text{sin}(\Omega_{SB_j}t+\delta_j)e^{-i\omega_{c_j}t}.
    \end{aligned}
\end{equation}
We substitute the drive term in \autoref{eq: System_H} and apply the first two steps of the process to get
\begin{equation}
\begin{aligned}
    \frac{H_2}{\hbar} = &\frac{\chi_j}{2} \hat{a_j}^\dagger \hat{a_j}\sigma_z + \frac{1}{2}[(\omega_q-\omega_d)\sigma_z + \Omega_R\sigma_x]\\
    &-i\alpha_j\Omega_{SB_j}\text{sin}(\Omega_{SB_j}t+\delta_j)(\hat{a_j}^\dagger-\hat{a_j}).
    \end{aligned}
\end{equation}
This drive generates a displacement in the cavity of the form. (See Appendix B1)
\begin{equation}
    \alpha_j(t)=\alpha_j\text{cos}(\Omega_{SB_j}t+\delta_j).
\end{equation}
We now perform a transformation into the displaced frame ($\hat{a_j} \rightarrow \hat{d_j} + \alpha_j$, where $\hat{a_j} = \hat{d_j}$, this new notation is used to avoid confusion), allowing for the elimination of the driving term in $H_2$ within this frame. (See Appendix B2)
\begin{equation}
\begin{aligned}
    \frac{H_3}{\hbar} &= \frac{\chi_j}{2}(\hat{d_j}^\dagger+\alpha_j^*(t))(\hat{d_j}+\alpha_j(t))\sigma_z \\& +\frac{1}{2}[(\omega_q-\omega_d)\sigma_z + \Omega_R\sigma_x] \\
    &= \frac{\chi_j}{2}[\alpha_j\text{cos}(\Omega_{SB_j}t+\delta_j)(\hat{d_j}^\dagger+\hat{d_j}) + \frac{\alpha_j^2}{2}\\
    &~~~~+ \frac{\alpha_j^2}{2}\text{cos}(2\Omega_{SB_j}t+2\delta_j) + \hat{d_j}^\dagger \hat{d_j}]\sigma_z\\
    &~~~~+ \frac{1}{2}[(\omega_q-\omega_d)\sigma_z + \Omega_R\sigma_x].
    \end{aligned}
\end{equation}
Choosing resonant Rabi drive by setting $\omega_q-\omega_d=-\chi\alpha_j^2/2$, we then apply step 5 of the transformation process,
\begin{equation}
\begin{aligned}
    \frac{H_5}{\hbar} &= (\frac{\chi_j\alpha_j}{2}\text{cos}(\Omega_{SB_j}t+\delta_j)(\hat{d_j}^\dagger+\hat{d_j})\\
    &+ \frac{\chi_j}{2}[\frac{\alpha_j^2}{2}\text{cos}(2\Omega_{SB_j}t+2\delta_j) + \hat{d_j}^\dagger \hat{d_j}])\\
    &\cdot{(\sigma e^{-i\Omega_{SB_j}t}+\sigma^\dagger e^{i\Omega_{SB_j}t})}\\
    &+ \frac{\Delta\Omega_j}{2}\sigma_z,
    \end{aligned}
\end{equation}
where $\Delta\Omega_j \equiv \Omega_R - \Omega_{SB_j}$, and apply the RWA.
\begin{equation}
    \frac{H_6}{\hbar} = \frac{\chi_j\alpha_j}{4}(\hat{d_j}^\dagger+\hat{d_j})(\sigma e^{i\delta_j}+\sigma^\dagger e^{-i\delta_j}) + \frac{\Delta\Omega_j}{2}\sigma_z.
\end{equation}
When we utilize the first cavity mode and $\delta_1=0$, we get the 1D Dirac Hamiltonian for free particle,
\begin{equation}
    \frac{H'_{1D}}{\hbar} = \frac{\chi_1\alpha_1}{4}(\hat{d_1}^\dagger+\hat{d_1})\sigma_x + \frac{\Delta\Omega_1}{2}\sigma_z.
\end{equation}
With 2 modes, when $\delta_1=0,~\delta_2=\pi/2$, we get the 2D Dirac Hamiltonian for free particle,
\begin{equation}
\begin{aligned}
    \frac{H'_{2D}}{\hbar} &= \frac{\chi_1\alpha_1}{4}(\hat{d_1}^\dagger+\hat{d_1})\sigma_x + \frac{\chi_2\alpha_2}{4}(\hat{d_2}^\dagger+\hat{d_2})\sigma_y \\ &+ \frac{\Delta\Omega_1+\Delta\Omega_2}{2}\sigma_z.
    \end{aligned}
\end{equation}
Hamiltonian mapping with Dirac Hamiltonian in \autoref{eq: Dirac-1D,2D},
\begin{equation}
\begin{aligned}
    &c\equiv\frac{\chi_1\alpha_1}{4}=\frac{\chi_2\alpha_2}{4}, ~mc^2\equiv \sum_j\frac{\Delta\Omega_j}{2},\\
    &\hat{P}_{x/y} \equiv \hat{d}_{1/2}^\dagger+\hat{d}_{1/2}.
    \end{aligned}
\end{equation}

\subsection{Dirac particle in a constant magnetic field} \label{app: magnetic}
To simulate the additional magnetic potential, we apply an asymmetric pair of sidebands, i.e. 
\begin{equation}
\begin{aligned}
    \epsilon_j(t) &= \Omega_{SB_j}(\frac{\alpha_j}{2}e^{-i(\Omega_{SB_j}t+\delta_j)}\\
    &~~~~~~~~~~~~~~~~~~~~-\frac{\alpha_j+\Delta\alpha_j}{2}e^{i(\Omega_{SB_j}t+\delta_j)})e^{-i\omega_{c_j}t}\\
    &= \Omega_{SB_j}(-\frac{\Delta\alpha_j}{2}e^{i(\Omega_{SB_j}t+\delta_j)}\\
    &~~~~~~~~~~~~~~~~~~~~-i\alpha_j\text{sin}(\Omega_{SB_j}t+\delta_j))e^{-i\omega_{c_j}t}.
    \end{aligned}
\end{equation}
The sum of the drives generates a displacement of the form
\begin{equation}
    \alpha_j(t)=\alpha_j\text{cos}(\Omega_{SB_j}t+\delta_j) + \frac{\Delta\alpha_j}{2}e^{i(\Omega_{SB_j}t+\delta_j)},
\end{equation}
which yields the displaced frame Hamiltonian,
\begin{equation}
\begin{aligned}
    \frac{H_3}{\hbar} &= \frac{\chi_j}{2}(\hat{d_j}^\dagger+\alpha_j^*(t))(\hat{d_j}+\alpha_j(t))\sigma_z \\& + \frac{1}{2}[(\omega_q-\omega_d)\sigma_z + \Omega_R\sigma_x]\\ 
    &= \frac{\chi_j}{2}[\alpha_j\text{cos}(\Omega_{SB_j}t+\delta_j)(\hat{d_j}^\dagger+\hat{d_j}) + \frac{\alpha_j^2}{2} \\
    & + \frac{\alpha_j^2}{2}\text{cos}(2\Omega_{SB_j}t+2\delta_j) + \frac{\alpha_j\Delta\alpha_j}{2}\\
    & + \frac{\alpha_j\Delta\alpha_j}{2}~\text{cos}(2\Omega_{SB_j}t+2\delta_j) + \frac{\Delta\alpha_j^2}{4}\\
    & + \frac{1}{2}(\hat{d_j}^\dagger\Delta\alpha_je^{i(\Omega_{SB_j}t+\delta_j)} + \hat{d_j}\Delta\alpha_je^{-i(\Omega_{SB_j}t+\delta_j)}) + \hat{d_j}^\dagger \hat{d_j}]\sigma_z\\
    & + \frac{1}{2}[(\omega_q-\omega_d)\sigma_z + \Omega_R\sigma_x].
    \end{aligned}
\end{equation}
Choosing resonant Rabi drive by setting $\omega_q-\omega_d=-\chi(\alpha_j^2/2+\alpha_j\Delta\alpha_j/2+\Delta\alpha_j^2/4)$, we then apply step 5 of the transformation process,
\begin{equation}
\begin{aligned}
    \frac{H_5}{\hbar} &= \frac{\chi_j}{2}[\alpha_j\text{cos}(\Omega_{SB_j}t+\delta_j)(\hat{d_j}^\dagger+\hat{d_j}) \\
    &+ \frac{\alpha_j^2}{2}\text{cos}(2\Omega_{SB_j}t+2\delta_j)\\
    &+ \frac{\alpha_j\Delta\alpha_j}{2}\text{cos}(2\Omega_{SB_j}t+2\delta_j)\\
    &+ \frac{1}{2}(\hat{d_j}^\dagger\Delta\alpha_je^{i(\Omega_{SB_j}t+\delta_j)} + \hat{d_j}\Delta\alpha_je^{-i(\Omega_{SB_j}t+\delta_j)}) + \hat{d_j}^\dagger \hat{d_j}]\\
    &\cdot(\sigma e^{-i\Omega_{SB_j}t}+\sigma^\dagger e^{i\Omega_{SB_j}t}) + \frac{\Delta\Omega_j}{2}\sigma_z,
    \end{aligned}
\end{equation}
and apply the RWA,
\begin{equation}
\begin{aligned}
    \frac{H_6}{\hbar} &= \frac{\chi_j\alpha_j}{4}(\hat{d_j}^\dagger+\hat{d_j})(\sigma e^{i\delta_j}+\sigma^\dagger e^{-i\delta_j})\\
    &+ \frac{\chi_j\Delta\alpha_j}{4}(\hat{d_j}^\dagger\sigma e^{i\delta_j} + \hat{d_j}\sigma^\dagger e^{-i\delta_j})\\
    &+ \frac{\Delta\Omega_j}{2}\sigma_z.
    \end{aligned}
\end{equation}
When we only use the first cavity mode, and $\delta_1=0$, we get the 1D Dirac Hamiltonian in a constant magnetic field,
\begin{equation}
\begin{aligned}
    \frac{H_{1D,\text{mag}}}{\hbar} &= \chi_1(\frac{\alpha_1}{4}+\frac{\Delta\alpha_1}{8})(\hat{d_1}^\dagger+\hat{d_1})\sigma_x \\
    &- \chi_1\frac{\Delta\alpha_1}{8}i(\hat{d_1}^\dagger-\hat{d_1})\sigma_y + \frac{\Delta\Omega_1}{2}\sigma_z.
    \end{aligned}
\end{equation}
By applying asymmetric sidebands to one mode and symmetric driving to the other, with the choices $\delta_1=0,~\delta_2=\pi/2$, we get the 2D Dirac Hamiltonian in a constant magnetic field,
\begin{equation}
\begin{aligned}
    \frac{H_{2D,\text{mag}}}{\hbar} &= \chi_1(\frac{\alpha_1}{4}+\frac{\Delta\alpha_1}{8})(\hat{d_1}^\dagger+\hat{d_1})\sigma_x\\ 
    &+\chi_2\frac{\alpha_2}{4}(\hat{d_2}^\dagger+\hat{d_2})\sigma_y\\
    &-\chi_1\frac{\Delta\alpha_1}{8}i(\hat{d_1}^\dagger-\hat{d_1})\sigma_y + \frac{\Delta\Omega_1+\Delta\Omega_2}{2}\sigma_z.
    \end{aligned}
\end{equation}
Hamiltonian mapping with Dirac Hamiltonian in \autoref{eq: Dirac in magnetic},
\begin{equation}
\begin{aligned}
    &c\equiv\chi_1(\frac{\alpha_1}{4}+\frac{\Delta\alpha_1}{8})=\chi_2\frac{\alpha_2}{4}, ~mc^2\equiv \sum_j\frac{\Delta\Omega_j}{2},\\
    &\hat{P}_{x/y}\equiv \hat{d}_{1/2}^\dagger+\hat{d}_{1/2}, ~\hat{X}/\hat{Y} \equiv \frac{i}{2}(\hat{d}_{1/2}-\hat{d}_{1/2}^\dagger),\\ 
    &eB\equiv \frac{2\Delta\alpha_1}{2\alpha_1+\Delta\alpha_1}.
    \end{aligned}
\end{equation}

\subsection{Dirac particle in an electrostatic potential} \label{app: elecrostatic}
Finally, to simulate an electrostatic potential we add a cavity mode resonant drive to a symmetric pair of sidebands, i.e.
\begin{equation}
\begin{aligned}
    \epsilon_j(t) = -i\alpha_j\Omega_{SB_j}\text{sin}(\Omega_{SB_j}t+\delta_j)e^{-i\omega_{c_j}t}+i\gamma_je^{-i\omega_{c_j}t}.
    \end{aligned}
\end{equation}
With this driving term, we apply the first two steps of the transformation process,
\begin{equation}
\begin{aligned}
    \frac{H_2}{\hbar} = &\frac{\chi_j}{2} \hat{a_j}^\dagger \hat{a_j}\sigma_z + \frac{1}{2}[(\omega_q-\omega_d)\sigma_z + \Omega_R\sigma_x]\\
    &-i\alpha_j\Omega_{SB_j}\text{sin}(\Omega_{SB_j}t+\delta_j)(\hat{a_j}^\dagger-\hat{a_j}) \\
    &+i\gamma_j(\hat{a_j}^\dagger-\hat{a_j}).
    \end{aligned}
\end{equation}
In the displaced cavity frame with $\alpha(t)=\alpha_j\text{cos}(\Omega_{SB_j}t+\delta_j)$,
\begin{equation}
\begin{aligned}
    \frac{H_3}{\hbar} &= \frac{\chi_j}{2}(\hat{d_j}^\dagger+\alpha_j^*(t))(\hat{d_j}+\alpha_j(t))\sigma_z \\
    &~~~~~~~+\frac{1}{2}[(\omega_q-\omega_d)\sigma_z + \Omega_R\sigma_x] \\
    &~~~~~~~+ i\gamma_j((\hat{d_j}^\dagger+\alpha_j^*(t))-(\hat{d_j}+\alpha_j(t)))\\
    &= \frac{\chi_j}{2}[\alpha_j\text{cos}(\Omega_{SB_j}t+\delta_j)(\hat{d_j}^\dagger+\hat{d_j}) + \frac{\alpha_j^2}{2}\\
    &~~~~~~~+ \frac{\alpha_j^2}{2}\text{cos}(2\Omega_{SB_j}t+2\delta_j) + \hat{d_j}^\dagger \hat{d_j}]\sigma_z\\
    &~~~~~~~+ \frac{1}{2}[(\omega_q-\omega_d)\sigma_z + \Omega_R\sigma_x] + \gamma_ji(\hat{d_j}^\dagger-\hat{d_j}).
    \end{aligned}
\end{equation}
Repeating exactly the last steps from \autoref{app: free} and choosing $\delta_1=0$, we get
\begin{equation}
    \frac{H'_\text{1D,elec}}{\hbar} = \frac{\chi_1\alpha_1}{4}(\hat{d_1}^\dagger+\hat{d_1})\sigma_x + \gamma_1i(\hat{d_1}^\dagger-\hat{d_1}) + \frac{\Delta\Omega_1}{2}\sigma_z,
\end{equation}
Hamiltonian mapping with Dirac Hamiltonian in \autoref{eq: Dirac-elec},
\begin{equation}
\begin{aligned}
    &c\equiv\chi_1\frac{\alpha_1}{4}, ~mc^2\equiv\frac{\Delta\Omega_1}{2}, ~ g\equiv -2\gamma_1,\\
    &\hat{P_x}\equiv \hat{d_1}^\dagger+\hat{d_1}, ~\hat{X} \equiv \frac{i}{2}(\hat{d_1}-\hat{d_1}^\dagger).
    \end{aligned}
\end{equation}

\section{Displaced frame} \label{app: displaced frame}

\subsection{Cavity state dynamics}
In this section, we will discuss the dynamics of the cavity state under the influence of sideband drivings. According to input-output theory, the Heisenberg equation of motion is described as follows:
\begin{equation} \label{eq: input-output}
    {\partial_t}\hat{a} = -\frac{i}{\hbar}[\hat{a},H_{total}]
\end{equation}
Here, $H_{total}$ represents the Hamiltonian that takes into account single cavity mode, the environment itself, and the interaction between the environment and the targeted cavity mode. In the cavity resonance frequency frame $\omega_c$, \autoref{eq: input-output} can be rewritten as follows.
\begin{equation}
    \dot{\alpha} = i\epsilon - \frac{\kappa}{2}\alpha
\end{equation}
where $\alpha = \langle\alpha|\hat{a}|\alpha\rangle$, $|\alpha\rangle$ is coherent state, $\epsilon = -i\alpha_0\Omega_{SB}\text{sin}(\Omega_{SB}t+\delta)$ ($\alpha_0\Omega_{SB}$ is arbitrary amplitude of sideband drives). Then, $\alpha(t)=\alpha_0\text{cos}(\Omega_{SB}t+\delta)$ in the cavity resonance frequency frame ($\kappa$ is negligible in high Q-factor cavity mode). This represents the cavity state oscillating at $\Omega_{SB}$ along the quadrature axes of cavity mode.

\subsection{Elimination of driving terms}

The master equation for a damped cavity, characterized by a decay rate $\kappa$, is given by:
\begin{equation}
    \frac{d\rho}{dt} = -\frac{i}{\hbar}[H,\rho] + \kappa D[\hat{a}]\rho
\end{equation}
where the dissipation operator is defined as $D[X]\rho = X\rho X^\dagger - \frac{X^\dagger X\rho + \rho X^\dagger X}{2}$, and $H$ represents the total Hamiltonian as described in \autoref{eq: System_H}. To eliminate the driving terms in $H$, we perform a transformation into the displaced cavity frame by applying the displacement operator $\mathcal{D}(\alpha(t)) = e^{(\alpha \hat{d}^\dagger - \alpha^* \hat{d})}$. The transformation proceeds as follows:

\begin{enumerate}
    \item The Hamiltonian transforms according to
    \begin{equation} 
    \begin{aligned}
            H \rightarrow H_{disp} &= 
            \mathcal{D}(\alpha)H\mathcal{D}^\dagger(\alpha)+i\frac{\partial \mathcal{D}(\alpha)}{\partial t}\mathcal{D}^\dagger(\alpha)\\ &= \mathcal{D}(\alpha)H\mathcal{D}^\dagger(\alpha) + H^{'}.
        \end{aligned}
    \end{equation}
    \item The dissipative term transforms as 
    \begin{equation}
        \kappa D[\hat{a}]\rho \rightarrow \kappa D[\hat{d}+\alpha]\rho = \kappa D[\hat{d}]\rho - i[H^{''}, \rho].
    \end{equation}
\end{enumerate}

Essentially, $\hat{a} = \hat{d}$; we are simply using a different symbol to avoid confusion. Thus, the modified master equation becomes:
\begin{equation}
    d\rho = -i[UHU^\dagger + H^{'},\rho]dt + \kappa D[\hat{d}]\rho dt - i[H^{''},\rho]dt,
\end{equation}
where $U = \mathcal{D}(\alpha)$, and
\begin{equation}
\begin{aligned}
    UHU^\dagger &= H_q + \epsilon(t)(\hat{d}^\dagger + \alpha^*) + \epsilon^*(t)(\hat{d}+\alpha)\\
    & + \frac{\chi}{2}(\hat{d}^\dagger + \alpha^*)(\hat{d} + \alpha)\sigma_z
    \end{aligned}.
\end{equation}
When we define $H_{drive} \equiv \epsilon(t)(\hat{d}^\dagger + \alpha^*) + \epsilon^*(t)(\hat{d}+\alpha)$ and the condition $H_{drive}+H^{'}+H^{''} = 0$ is satisfied,
\begin{equation}
    \epsilon(t) = -i\alpha\sqrt{\Omega_{SB}^2+\frac{\kappa^2}{4}}\text{sin}(\Omega_{SB}t+\delta-\text{arctan}(\frac{\kappa}{2\Omega_{SB}})).
\end{equation}
Therefore, the master equation in the displaced frame is expressed as
\begin{equation}
    d\rho = -i[H_q+\frac{\chi}{2}(\hat{d}^\dagger + \alpha^*)(\hat{d} + \alpha)\sigma_z,\rho]dt + \kappa D[\hat{d}]\rho dt
\end{equation}
This equation represents the master equation transformed into the displaced frame, wherein the driving terms have been effectively eliminated.

\section{Beyond the RWA} \label{app: RWA}

The divergence between the Dirac Hamiltonian and the Dirac-like model Hamiltonian derived in circuit QED becomes more prominent at longer times. This divergence is primarily attributed to the term oscillating at $\Omega_{SB}$ in the sidebands driving frame,
\begin{equation}
    H_{\text{osc}}= \frac{\chi}{2}\hat{d}^\dagger \hat{d}\left(\sigma^\dagger e^{i\Omega_{SB}t}+\sigma e^{-i\Omega_{SB}t}\right).
\end{equation}
In general, the term oscillating at $\Omega_{SB}$ can be neglected. However, over longer times, the photon number in the cavity accumulates, making the number operator $\hat{d}^\dagger \hat{d}$ more significant. We will calculate the second-order effect of $H_{\text{osc}}$ using the Magnus expansion.

\begin{equation}
\begin{aligned}
    &H^{(2)}_{\text{Magnus}} \\
    &=\frac{-i}{2t_c} \int_{t}^{t+t_c}d\tau'\int_{t}^{\tau'}d\tau \left[H_{\text{osc}}(\tau '),H_{\text{osc}}(\tau)\right] \\ 
    &= \frac{-i\chi^2}{8t_c}(\hat{d}^\dagger \hat{d})^2\int_{t}^{t+t_c}d\tau'\int_{t}^{\tau'}d\tau \left[2i\sigma_z\sin\left(\Omega_{SB}\left(\tau '-\tau\right)\right)\right] \\
    &= \frac{\chi^2}{4t_c}(\hat{d}^\dagger \hat{d})^2\sigma_z \int_{t}^{t+t_c}d\tau'\int_{t}^{\tau'}d\tau \left[\sin\left(\Omega_{SB}\left(\tau '-\tau\right)\right)\right] \\
    &=-\frac{\chi^2}{4t_c\Omega_{SB}}(\hat{d}^\dagger \hat{d})^2\sigma_z \int_{t}^{t+t_c}d\tau' \left[\cos\left(\Omega_{SB}\left(\tau '-\tau\right)\right)\right]_t^{\tau '} \\
     &=-\frac{\chi^2}{4t_c\Omega_{SB}}(\hat{d}^\dagger \hat{d})^2\sigma_z \int_{t}^{t+t_c}d\tau' \left(1-\cos\left(\Omega_{SB}\left(\tau '-t\right)\right)\right) \\
     &=\frac{\chi^2}{4t_c\Omega_{SB}}(\hat{d}^\dagger \hat{d})^2\sigma_z \left(t_c-\left[\frac{1}{\Omega_{SB}}\sin\left(\Omega_{SB}\left(\tau '-t\right)\right)\right]_t^{t+t_c}\right) \\
     &=\frac{\chi^2}{4t_c\Omega_{SB}}(\hat{d}^\dagger \hat{d})^2\sigma_z \left(t_c-\frac{1}{\Omega_{SB}}\sin\left(\Omega_{SB}\left(t_c\right)\right)\right) \\
     &=\frac{\chi^2}{8\pi\Omega_{SB}}(\hat{d}^\dagger \hat{d})^2\sigma_z \left(2\pi-\sin\left(\Omega_{SB}\left(t_c\right)\right)\right) \\
     &\approx \frac{\chi^2}{4\Omega_{SB}}(\hat{d}^\dagger \hat{d})^2\sigma_z,
\end{aligned}
\end{equation}
where the commutation relation reads
\begin{equation}
\begin{aligned}
    &\left[H_{\text{osc}}(\tau '),H_{\text{osc}}(\tau)\right] \\
    &=\left[\sigma^\dagger e^{i\Omega_{SB}\tau '}+\sigma e^{-i\Omega_{SB}\tau '},\sigma^\dagger e^{i\Omega_{SB}\tau}+\sigma e^{-i\Omega_{SB}\tau}\right] \\
    &=\left[\sigma^\dagger ,\sigma \right]\left(e^{i\Omega_{SB}\left(\tau '-\tau\right)} - e^{-i\Omega_{SB}\left(\tau '-\tau\right)}\right)\\
    &=2i\sigma_z\sin\left(\Omega_{SB}\left(\tau '-\tau\right)\right).
\end{aligned}
\end{equation}
The second order correction to the Hamiltonian gives a photon number dependent shift in the mass term which scales with $\frac{\chi^2n^2}{4\Omega_{SB}}$, where $n$ is the number of photons in the cavity.

\begin{figure}
    \centering
    \includegraphics[width=0.4\textwidth]{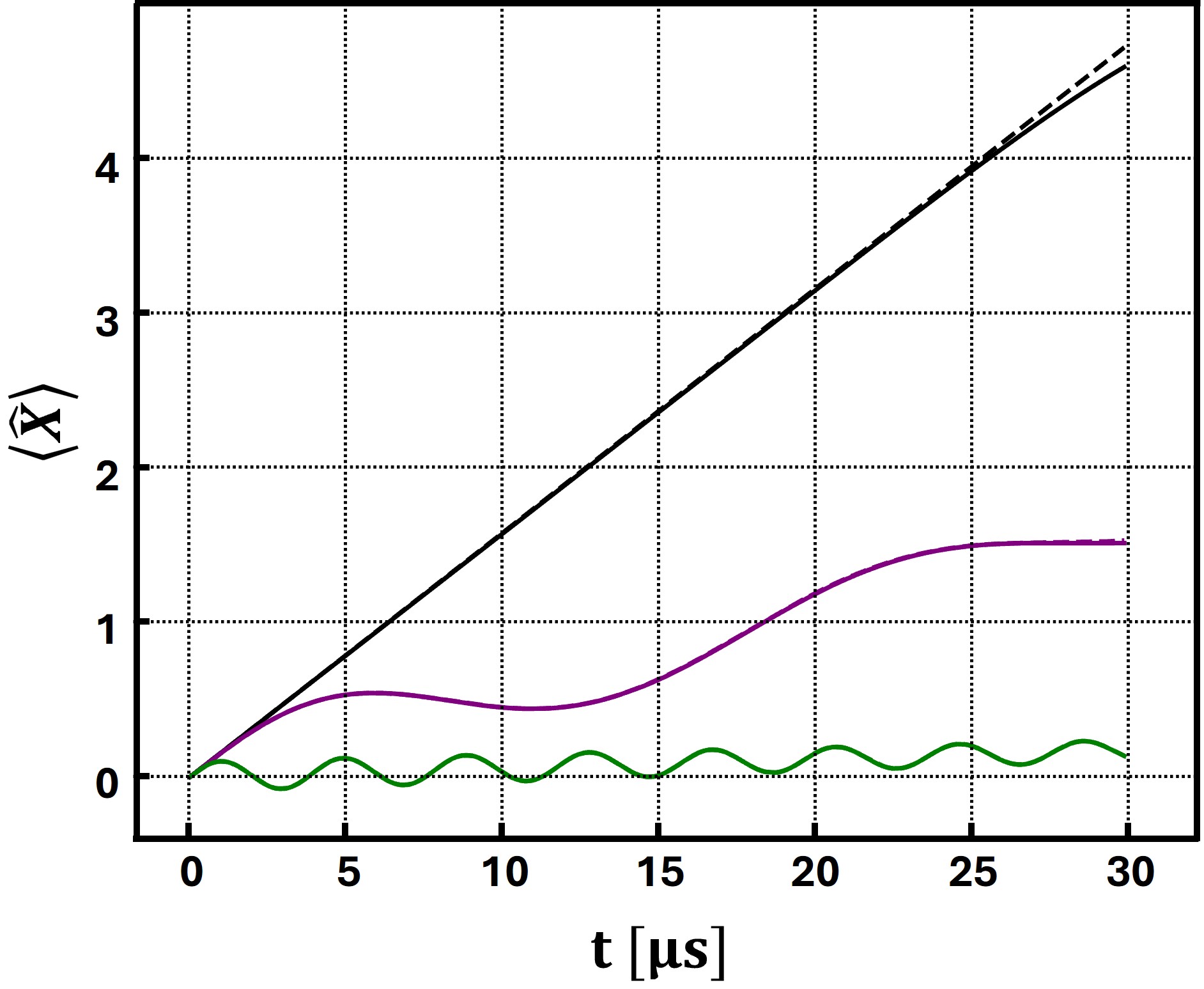}
    \caption{Zitterbewegung simulations of a 1D free Dirac particle with a sideband frequency set to 100 MHz. In the main text figures, a sideband frequency of 40 MHz was used. The larger sideband frequency in this case demonstrates that the rotating wave approximation(RWA) remains valid for a longer time evolution.}
    \label{fig:appendix-c}
\end{figure}

\section{Discrete Landau energy levels in Fourier transform of qubit $ \langle \sigma_z(t) \rangle $}

The Landau energy levels of a Dirac particle can be determined by performing the Fourier transform of the time-dependent expectation value of the qubit's $\langle\sigma_z(t)\rangle$. By analyzing the frequency components obtained from the Fourier transform, one can extract the energy level differences corresponding to Landau quantization. This theoretical part is referenced by supplementary materials of \cite{Jiang2022}.

We will explain how the calculation of the Landau energy levels derived from the actual Weyl particle Hamiltonian corresponds to the calculations of our system's Hamiltonian. As shown in \cite{Jiang2022}, The Landau energy levels derived from the 2D Weyl Hamiltonian in a magnetic field are identical to those obtained from the 1D Weyl Hamiltonian. Therefore, the discussion can begin with the 1D Weyl Hamiltonian in the presence of a magnetic field.
\begin{equation} \label{eq: 1D magnetic Weyl}
    H_{\text{1D,Weyl,mag}} = c\sigma_x\hat{P_x}-ceB\hat{X}\sigma_y.
\end{equation}

In our system, the emulated 1D Weyl Hamiltonian from asymmetric sideband driving is
\begin{equation}
    H/\hbar = (\frac{\chi_1{\alpha_1}}{4}+\frac{\chi_1{\Delta\alpha_1}}{8})(\hat{d_1}^{\dagger}+\hat{d_1})\sigma_x - \frac{i\chi_1\Delta\alpha_1}{8}(\hat{d_1}^{\dagger}-\hat{d_1})\sigma_y.
\end{equation}

In this case, a different Hamiltonian mapping method is used, distinct from the one based on Equation A19. By setting $c=1$, the quadratures are defined as $\hat{P_x} = C(\hat{d_1}^{\dagger}+\hat{d_1})$, $\hat{X} = \frac{i}{2C}(\hat{d_1}-\hat{d_1}^{\dagger})$ which satisfy the commutation relation $[\hat{X},\hat{P_x}] = i$, where $C=\frac{\chi_1{\alpha_1}}{4}+\frac{\chi_1{\Delta\alpha_1}}{8}$, $\frac{eB}{2C} = -\frac{\chi_1{\Delta\alpha_1}}{8}$.

If we utilize new Hamiltonian mapping method mentioned in above, \autoref{eq: 1D magnetic Weyl} can be expressed as 
\begin{equation}
\begin{aligned}
    H_{\text{1D,Weyl,mag}} &= C(\hat{d_1}^\dagger+\hat{d_1})\sigma_x - \frac{eB}{2C}i(\hat{d_1}-\hat{d_1}^\dagger)\sigma_y\\
    &= C(\hat{d_1}^\dagger+\hat{d_1})(\sigma_+ + \sigma_-)\\ 
    &~~~~~+ \frac{eB}{2C}(\hat{d_1}-\hat{d_1}^\dagger)(\sigma_+ - \sigma_-)\\
    &= (C-\frac{eB}{2C})(\hat{d_1}^{\dagger}\sigma_+ + \hat{d_1}\sigma_-) \\
    &~~~~~+ (C + \frac{eB}{2C})(\hat{d_1}^{\dagger}\sigma_- + \hat{d_1}\sigma_+).
\end{aligned}
\end{equation}
When we set $\alpha_1 = -\Delta\alpha_1$ which means $C-\frac{eB}{2C}=0$,
\begin{equation}
\begin{aligned}
    H_{\text{1D,Weyl,mag}} &= (C + \frac{eB}{2C})(\hat{d_1}^{\dagger}\sigma_- + \hat{d_1}\sigma_+) \\
    &= \frac{\chi_1\alpha_1}{4}(\hat{d_1}^{\dagger}\sigma_- + \hat{d_1}\sigma_+)
\end{aligned}
\end{equation}
This Hamiltonian is mapped onto $H = \sqrt{2eB}(\hat{d_1}^\dagger\sigma_- + \hat{d_1}\sigma_+)$ from \cite{Jiang2022}, establishing that $\sqrt{2eB}$ corresponds to $\frac{\chi_1\alpha_1}{4}$.
Therefore, the quantized energy level difference is $E_n^+ - E_n^- = 2\sqrt{2neB} = \frac{\chi_1}{2}\alpha_1\sqrt{n}$, where $E_n^+$, $E_n^-$ are eigenvalues of $H$.

In the main text, we simulated the quantized energy level differences using the parameters $\alpha_1 = 1$, $\Delta\alpha_1 = -1$, $\frac{\chi_1}{2\pi} = 0.1\text{MHz}$, and $\Delta\Omega_1 = 0$ (Weyl particle).  A total evolution time of 5000$\mu s$ was chosen to achieve sufficient resolution in the Fourier transform results, although higher peaks remain difficult to distinguish. Under these conditions, the first peak was observed at 0.05 MHz. Additionally, we simulated the massive Dirac particle case with $\frac{\Delta\Omega_1}{2\pi} = 0.05\text{MHz}$. In this scenario, \autoref{eq: 1D magnetic Weyl} is modified into 
\begin{equation}
    H_{\text{1D,Dirac,mag}} = \frac{\chi_1\alpha_1}{4}(\hat{d_1}^\dagger\sigma_- + \hat{d_1}\sigma_+) + \frac{\Delta\Omega_1}{2}\sigma_z.
\end{equation}
Therefore, the Landau energy difference becomes $E_n^+ - E_n^- = 2\sqrt{2neB + m^2} = 2\sqrt{(\frac{\chi_1\alpha_1}{4})^2n + (\frac{\Delta\Omega_1}{2})^2}$.

\begin{figure}
    \centering
    \includegraphics[width=0.4\textwidth]{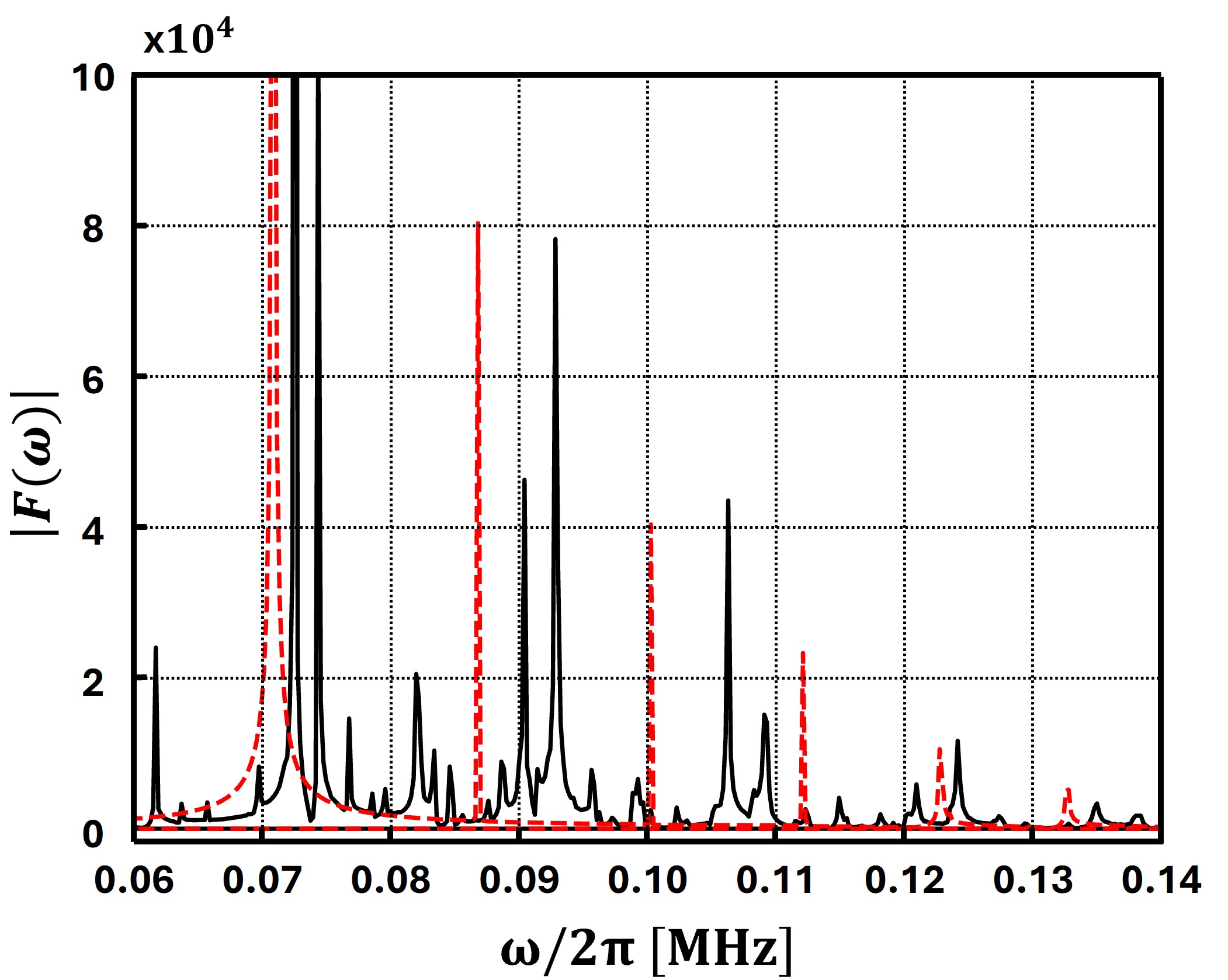}
    \caption{Fourier transform result of time-dependent expectation value of qubit's $\langle\sigma_z(t)\rangle$ for Dirac (massive) particle in magnetic field. It simulated with same conditions in main text except mass term ($\frac{\Delta\Omega_1}{2\pi} = 0.05\text{MHz}$). Thus, we can find the first peak at $E_1^+ - E_1^- = 2\sqrt{2eB + m^2}. = 0.0707\text{MHz}$.}
    \label{fig: dirac_energy_level}
\end{figure}

\clearpage

\bibliography{apssamp}

\begin{thebibliography}{46}%
\makeatletter
\providecommand \@ifxundefined [1]{%
 \@ifx{#1\undefined}
}%
\providecommand \@ifnum [1]{%
 \ifnum #1\expandafter \@firstoftwo
 \else \expandafter \@secondoftwo
 \fi
}%
\providecommand \@ifx [1]{%
 \ifx #1\expandafter \@firstoftwo
 \else \expandafter \@secondoftwo
 \fi
}%
\providecommand \natexlab [1]{#1}%
\providecommand \enquote  [1]{``#1''}%
\providecommand \bibnamefont  [1]{#1}%
\providecommand \bibfnamefont [1]{#1}%
\providecommand \citenamefont [1]{#1}%
\providecommand \href@noop [0]{\@secondoftwo}%
\providecommand \href [0]{\begingroup \@sanitize@url \@href}%
\providecommand \@href[1]{\@@startlink{#1}\@@href}%
\providecommand \@@href[1]{\endgroup#1\@@endlink}%
\providecommand \@sanitize@url [0]{\catcode `\\12\catcode `\$12\catcode
  `\&12\catcode `\#12\catcode `\^12\catcode `\_12\catcode `\%12\relax}%
\providecommand \@@startlink[1]{}%
\providecommand \@@endlink[0]{}%
\providecommand \url  [0]{\begingroup\@sanitize@url \@url }%
\providecommand \@url [1]{\endgroup\@href {#1}{\urlprefix }}%
\providecommand \urlprefix  [0]{URL }%
\providecommand \Eprint [0]{\href }%
\providecommand \doibase [0]{https://doi.org/}%
\providecommand \selectlanguage [0]{\@gobble}%
\providecommand \bibinfo  [0]{\@secondoftwo}%
\providecommand \bibfield  [0]{\@secondoftwo}%
\providecommand \translation [1]{[#1]}%
\providecommand \BibitemOpen [0]{}%
\providecommand \bibitemStop [0]{}%
\providecommand \bibitemNoStop [0]{.\EOS\space}%
\providecommand \EOS [0]{\spacefactor3000\relax}%
\providecommand \BibitemShut  [1]{\csname bibitem#1\endcsname}%
\let\auto@bib@innerbib\@empty
\bibitem [{\citenamefont {Feynman}(1982)}]{Feynman1982}%
  \BibitemOpen
  \bibfield  {author} {\bibinfo {author} {\bibfnamefont {R.~P.}\ \bibnamefont
  {Feynman}},\ }\bibfield  {title} {\bibinfo {title} {Simulating physics with
  computers},\ }\href@noop {} {\bibfield  {journal} {\bibinfo  {journal}
  {Inernational Journal of Theoretical Physics}\ }\textbf {\bibinfo {volume}
  {21}} (\bibinfo {year} {1982})}\BibitemShut {NoStop}%
\bibitem [{\citenamefont {Buluta}\ and\ \citenamefont
  {Nori}(2009)}]{Buluta2009}%
  \BibitemOpen
  \bibfield  {author} {\bibinfo {author} {\bibfnamefont {I.}~\bibnamefont
  {Buluta}}\ and\ \bibinfo {author} {\bibfnamefont {F.}~\bibnamefont {Nori}},\
  }\bibfield  {title} {\bibinfo {title} {Quantum simulators},\ }\href
  {https://doi.org/10.1126/science.1177838} {\bibfield  {journal} {\bibinfo
  {journal} {Science}\ }\textbf {\bibinfo {volume} {326}},\ \bibinfo {pages}
  {108} (\bibinfo {year} {2009})},\ \Eprint
  {https://arxiv.org/abs/https://www.science.org/doi/pdf/10.1126/science.1177838}
  {https://www.science.org/doi/pdf/10.1126/science.1177838} \BibitemShut
  {NoStop}%
\bibitem [{\citenamefont {Friedenauer}\ \emph {et~al.}(2008)\citenamefont
  {Friedenauer}, \citenamefont {Schmitz}, \citenamefont {Glueckert},
  \citenamefont {Porras},\ and\ \citenamefont {Schaetz}}]{Friedenauer2008}%
  \BibitemOpen
  \bibfield  {author} {\bibinfo {author} {\bibfnamefont {A.}~\bibnamefont
  {Friedenauer}}, \bibinfo {author} {\bibfnamefont {H.}~\bibnamefont
  {Schmitz}}, \bibinfo {author} {\bibfnamefont {J.~T.}\ \bibnamefont
  {Glueckert}}, \bibinfo {author} {\bibfnamefont {D.}~\bibnamefont {Porras}},\
  and\ \bibinfo {author} {\bibfnamefont {T.}~\bibnamefont {Schaetz}},\
  }\bibfield  {title} {\bibinfo {title} {Simulating a quantum magnet with
  trapped ions},\ }\href {https://doi.org/10.1038/nphys1032} {\bibfield
  {journal} {\bibinfo  {journal} {Nature Physics}\ }\textbf {\bibinfo {volume}
  {4}},\ \bibinfo {pages} {757} (\bibinfo {year} {2008})}\BibitemShut {NoStop}%
\bibitem [{\citenamefont {Lamata}\ \emph {et~al.}(2007)\citenamefont {Lamata},
  \citenamefont {Le{\'o}n}, \citenamefont {Sch{\"a}tz},\ and\ \citenamefont
  {Solano}}]{Lamata2007}%
  \BibitemOpen
  \bibfield  {author} {\bibinfo {author} {\bibfnamefont {L.}~\bibnamefont
  {Lamata}}, \bibinfo {author} {\bibfnamefont {J.}~\bibnamefont {Le{\'o}n}},
  \bibinfo {author} {\bibfnamefont {T.}~\bibnamefont {Sch{\"a}tz}},\ and\
  \bibinfo {author} {\bibfnamefont {E.}~\bibnamefont {Solano}},\ }\bibfield
  {title} {\bibinfo {title} {Dirac equation and quantum relativistic effects in
  a single trapped ion},\ }\href@noop {} {\bibfield  {journal} {\bibinfo
  {journal} {Physical review letters}\ }\textbf {\bibinfo {volume} {98}},\
  \bibinfo {pages} {253005} (\bibinfo {year} {2007})}\BibitemShut {NoStop}%
\bibitem [{\citenamefont {Flurin}\ \emph
  {et~al.}(2017{\natexlab{a}})\citenamefont {Flurin}, \citenamefont {Ramasesh},
  \citenamefont {Hacohen-Gourgy}, \citenamefont {Martin}, \citenamefont {Yao},\
  and\ \citenamefont {Siddiqi}}]{PhysRevX.7.031023}%
  \BibitemOpen
  \bibfield  {author} {\bibinfo {author} {\bibfnamefont {E.}~\bibnamefont
  {Flurin}}, \bibinfo {author} {\bibfnamefont {V.~V.}\ \bibnamefont
  {Ramasesh}}, \bibinfo {author} {\bibfnamefont {S.}~\bibnamefont
  {Hacohen-Gourgy}}, \bibinfo {author} {\bibfnamefont {L.~S.}\ \bibnamefont
  {Martin}}, \bibinfo {author} {\bibfnamefont {N.~Y.}\ \bibnamefont {Yao}},\
  and\ \bibinfo {author} {\bibfnamefont {I.}~\bibnamefont {Siddiqi}},\
  }\bibfield  {title} {\bibinfo {title} {Observing topological invariants using
  quantum walks in superconducting circuits},\ }\href
  {https://doi.org/10.1103/PhysRevX.7.031023} {\bibfield  {journal} {\bibinfo
  {journal} {Phys. Rev. X}\ }\textbf {\bibinfo {volume} {7}},\ \bibinfo {pages}
  {031023} (\bibinfo {year} {2017}{\natexlab{a}})}\BibitemShut {NoStop}%
\bibitem [{\citenamefont {Aspuru-Guzik}\ \emph {et~al.}(2005)\citenamefont
  {Aspuru-Guzik}, \citenamefont {Dutoi}, \citenamefont {Love},\ and\
  \citenamefont {Head-Gordon}}]{doi:10.1126/science.1113479}%
  \BibitemOpen
  \bibfield  {author} {\bibinfo {author} {\bibfnamefont {A.}~\bibnamefont
  {Aspuru-Guzik}}, \bibinfo {author} {\bibfnamefont {A.~D.}\ \bibnamefont
  {Dutoi}}, \bibinfo {author} {\bibfnamefont {P.~J.}\ \bibnamefont {Love}},\
  and\ \bibinfo {author} {\bibfnamefont {M.}~\bibnamefont {Head-Gordon}},\
  }\bibfield  {title} {\bibinfo {title} {Simulated quantum computation of
  molecular energies},\ }\href {https://doi.org/10.1126/science.1113479}
  {\bibfield  {journal} {\bibinfo  {journal} {Science}\ }\textbf {\bibinfo
  {volume} {309}},\ \bibinfo {pages} {1704} (\bibinfo {year} {2005})},\ \Eprint
  {https://arxiv.org/abs/https://www.science.org/doi/pdf/10.1126/science.1113479}
  {https://www.science.org/doi/pdf/10.1126/science.1113479} \BibitemShut
  {NoStop}%
\bibitem [{\citenamefont {Georgescu}\ \emph {et~al.}(2014)\citenamefont
  {Georgescu}, \citenamefont {Ashhab},\ and\ \citenamefont
  {Nori}}]{RevModPhys.86.153}%
  \BibitemOpen
  \bibfield  {author} {\bibinfo {author} {\bibfnamefont {I.~M.}\ \bibnamefont
  {Georgescu}}, \bibinfo {author} {\bibfnamefont {S.}~\bibnamefont {Ashhab}},\
  and\ \bibinfo {author} {\bibfnamefont {F.}~\bibnamefont {Nori}},\ }\bibfield
  {title} {\bibinfo {title} {Quantum simulation},\ }\href
  {https://doi.org/10.1103/RevModPhys.86.153} {\bibfield  {journal} {\bibinfo
  {journal} {Rev. Mod. Phys.}\ }\textbf {\bibinfo {volume} {86}},\ \bibinfo
  {pages} {153} (\bibinfo {year} {2014})}\BibitemShut {NoStop}%
\bibitem [{\citenamefont {Bloch}\ \emph {et~al.}(2012)\citenamefont {Bloch},
  \citenamefont {Dalibard},\ and\ \citenamefont {Nascimb{\`e}ne}}]{Bloch2012}%
  \BibitemOpen
  \bibfield  {author} {\bibinfo {author} {\bibfnamefont {I.}~\bibnamefont
  {Bloch}}, \bibinfo {author} {\bibfnamefont {J.}~\bibnamefont {Dalibard}},\
  and\ \bibinfo {author} {\bibfnamefont {S.}~\bibnamefont {Nascimb{\`e}ne}},\
  }\bibfield  {title} {\bibinfo {title} {Quantum simulations with ultracold
  quantum gases},\ }\href {https://doi.org/10.1038/nphys2259} {\bibfield
  {journal} {\bibinfo  {journal} {Nature Physics}\ }\textbf {\bibinfo {volume}
  {8}},\ \bibinfo {pages} {267} (\bibinfo {year} {2012})}\BibitemShut {NoStop}%
\bibitem [{\citenamefont {Blatt}\ and\ \citenamefont {Roos}(2012)}]{Blatt2012}%
  \BibitemOpen
  \bibfield  {author} {\bibinfo {author} {\bibfnamefont {R.}~\bibnamefont
  {Blatt}}\ and\ \bibinfo {author} {\bibfnamefont {C.~F.}\ \bibnamefont
  {Roos}},\ }\bibfield  {title} {\bibinfo {title} {Quantum simulations with
  trapped ions},\ }\href {https://doi.org/10.1038/nphys2252} {\bibfield
  {journal} {\bibinfo  {journal} {Nature Physics}\ }\textbf {\bibinfo {volume}
  {8}},\ \bibinfo {pages} {277} (\bibinfo {year} {2012})}\BibitemShut {NoStop}%
\bibitem [{\citenamefont {Aspuru-Guzik}\ and\ \citenamefont
  {Walther}(2012)}]{Aspuru-Guzik2012}%
  \BibitemOpen
  \bibfield  {author} {\bibinfo {author} {\bibfnamefont {A.}~\bibnamefont
  {Aspuru-Guzik}}\ and\ \bibinfo {author} {\bibfnamefont {P.}~\bibnamefont
  {Walther}},\ }\bibfield  {title} {\bibinfo {title} {Photonic quantum
  simulators},\ }\href {https://doi.org/10.1038/nphys2253} {\bibfield
  {journal} {\bibinfo  {journal} {Nature Physics}\ }\textbf {\bibinfo {volume}
  {8}},\ \bibinfo {pages} {285} (\bibinfo {year} {2012})}\BibitemShut {NoStop}%
\bibitem [{\citenamefont {Hensgens}\ \emph {et~al.}(2017)\citenamefont
  {Hensgens}, \citenamefont {Fujita}, \citenamefont {Janssen}, \citenamefont
  {Li}, \citenamefont {Van~Diepen}, \citenamefont {Reichl}, \citenamefont
  {Wegscheider}, \citenamefont {Das~Sarma},\ and\ \citenamefont
  {Vandersypen}}]{Hensgens2017}%
  \BibitemOpen
  \bibfield  {author} {\bibinfo {author} {\bibfnamefont {T.}~\bibnamefont
  {Hensgens}}, \bibinfo {author} {\bibfnamefont {T.}~\bibnamefont {Fujita}},
  \bibinfo {author} {\bibfnamefont {L.}~\bibnamefont {Janssen}}, \bibinfo
  {author} {\bibfnamefont {X.}~\bibnamefont {Li}}, \bibinfo {author}
  {\bibfnamefont {C.~J.}\ \bibnamefont {Van~Diepen}}, \bibinfo {author}
  {\bibfnamefont {C.}~\bibnamefont {Reichl}}, \bibinfo {author} {\bibfnamefont
  {W.}~\bibnamefont {Wegscheider}}, \bibinfo {author} {\bibfnamefont
  {S.}~\bibnamefont {Das~Sarma}},\ and\ \bibinfo {author} {\bibfnamefont
  {L.~M.~K.}\ \bibnamefont {Vandersypen}},\ }\bibfield  {title} {\bibinfo
  {title} {Quantum simulation of a fermi--hubbard model using a semiconductor
  quantum dot array},\ }\href {https://doi.org/10.1038/nature23022} {\bibfield
  {journal} {\bibinfo  {journal} {Nature}\ }\textbf {\bibinfo {volume} {548}},\
  \bibinfo {pages} {70} (\bibinfo {year} {2017})}\BibitemShut {NoStop}%
\bibitem [{\citenamefont {Houck}\ \emph {et~al.}(2012)\citenamefont {Houck},
  \citenamefont {T{\"u}reci},\ and\ \citenamefont {Koch}}]{Houck2012}%
  \BibitemOpen
  \bibfield  {author} {\bibinfo {author} {\bibfnamefont {A.~A.}\ \bibnamefont
  {Houck}}, \bibinfo {author} {\bibfnamefont {H.~E.}\ \bibnamefont
  {T{\"u}reci}},\ and\ \bibinfo {author} {\bibfnamefont {J.}~\bibnamefont
  {Koch}},\ }\bibfield  {title} {\bibinfo {title} {On-chip quantum simulation
  with superconducting circuits},\ }\href {https://doi.org/10.1038/nphys2251}
  {\bibfield  {journal} {\bibinfo  {journal} {Nature Physics}\ }\textbf
  {\bibinfo {volume} {8}},\ \bibinfo {pages} {292} (\bibinfo {year}
  {2012})}\BibitemShut {NoStop}%
\bibitem [{\citenamefont {Lamata}\ \emph {et~al.}(2018)\citenamefont {Lamata},
  \citenamefont {Parra-Rodriguez}, \citenamefont {Sanz},\ and\ \citenamefont
  {Solano}}]{Lamata_2018}%
  \BibitemOpen
  \bibfield  {author} {\bibinfo {author} {\bibfnamefont {L.}~\bibnamefont
  {Lamata}}, \bibinfo {author} {\bibfnamefont {A.}~\bibnamefont
  {Parra-Rodriguez}}, \bibinfo {author} {\bibfnamefont {M.}~\bibnamefont
  {Sanz}},\ and\ \bibinfo {author} {\bibfnamefont {E.}~\bibnamefont {Solano}},\
  }\bibfield  {title} {\bibinfo {title} {Digital-analog quantum simulations
  with superconducting circuits},\ }\href
  {https://doi.org/10.1080/23746149.2018.1457981} {\bibfield  {journal}
  {\bibinfo  {journal} {Advances in Physics: X}\ }\textbf {\bibinfo {volume}
  {3}},\ \bibinfo {pages} {1457981} (\bibinfo {year} {2018})}\BibitemShut
  {NoStop}%
\bibitem [{\citenamefont {Blais}\ \emph {et~al.}(2004)\citenamefont {Blais},
  \citenamefont {Huang}, \citenamefont {Wallraff}, \citenamefont {Girvin},\
  and\ \citenamefont {Schoelkopf}}]{PhysRevA.69.062320}%
  \BibitemOpen
  \bibfield  {author} {\bibinfo {author} {\bibfnamefont {A.}~\bibnamefont
  {Blais}}, \bibinfo {author} {\bibfnamefont {R.-S.}\ \bibnamefont {Huang}},
  \bibinfo {author} {\bibfnamefont {A.}~\bibnamefont {Wallraff}}, \bibinfo
  {author} {\bibfnamefont {S.~M.}\ \bibnamefont {Girvin}},\ and\ \bibinfo
  {author} {\bibfnamefont {R.~J.}\ \bibnamefont {Schoelkopf}},\ }\bibfield
  {title} {\bibinfo {title} {Cavity quantum electrodynamics for superconducting
  electrical circuits: An architecture for quantum computation},\ }\href
  {https://doi.org/10.1103/PhysRevA.69.062320} {\bibfield  {journal} {\bibinfo
  {journal} {Phys. Rev. A}\ }\textbf {\bibinfo {volume} {69}},\ \bibinfo
  {pages} {062320} (\bibinfo {year} {2004})}\BibitemShut {NoStop}%
\bibitem [{\citenamefont {Koch}\ \emph {et~al.}(2007)\citenamefont {Koch},
  \citenamefont {Yu}, \citenamefont {Gambetta}, \citenamefont {Houck},
  \citenamefont {Schuster}, \citenamefont {Majer}, \citenamefont {Blais},
  \citenamefont {Devoret}, \citenamefont {Girvin},\ and\ \citenamefont
  {Schoelkopf}}]{TransmonPaper}%
  \BibitemOpen
  \bibfield  {author} {\bibinfo {author} {\bibfnamefont {J.}~\bibnamefont
  {Koch}}, \bibinfo {author} {\bibfnamefont {T.~M.}\ \bibnamefont {Yu}},
  \bibinfo {author} {\bibfnamefont {J.}~\bibnamefont {Gambetta}}, \bibinfo
  {author} {\bibfnamefont {A.~A.}\ \bibnamefont {Houck}}, \bibinfo {author}
  {\bibfnamefont {D.~I.}\ \bibnamefont {Schuster}}, \bibinfo {author}
  {\bibfnamefont {J.}~\bibnamefont {Majer}}, \bibinfo {author} {\bibfnamefont
  {A.}~\bibnamefont {Blais}}, \bibinfo {author} {\bibfnamefont {M.~H.}\
  \bibnamefont {Devoret}}, \bibinfo {author} {\bibfnamefont {S.~M.}\
  \bibnamefont {Girvin}},\ and\ \bibinfo {author} {\bibfnamefont {R.~J.}\
  \bibnamefont {Schoelkopf}},\ }\bibfield  {title} {\bibinfo {title}
  {Charge-insensitive qubit design derived from the cooper pair box},\ }\href
  {https://doi.org/10.1103/PhysRevA.76.042319} {\bibfield  {journal} {\bibinfo
  {journal} {Phys. Rev. A}\ }\textbf {\bibinfo {volume} {76}},\ \bibinfo
  {pages} {042319} (\bibinfo {year} {2007})}\BibitemShut {NoStop}%
\bibitem [{\citenamefont {Kjaergaard}\ \emph {et~al.}(2020)\citenamefont
  {Kjaergaard}, \citenamefont {Schwartz}, \citenamefont {Braumüller},
  \citenamefont {Krantz}, \citenamefont {Wang}, \citenamefont {Gustavsson},\
  and\ \citenamefont {Oliver}}]{Kjaergaard2020}%
  \BibitemOpen
  \bibfield  {author} {\bibinfo {author} {\bibfnamefont {M.}~\bibnamefont
  {Kjaergaard}}, \bibinfo {author} {\bibfnamefont {M.~E.}\ \bibnamefont
  {Schwartz}}, \bibinfo {author} {\bibfnamefont {J.}~\bibnamefont
  {Braumüller}}, \bibinfo {author} {\bibfnamefont {P.}~\bibnamefont {Krantz}},
  \bibinfo {author} {\bibfnamefont {J.~I.-J.}\ \bibnamefont {Wang}}, \bibinfo
  {author} {\bibfnamefont {S.}~\bibnamefont {Gustavsson}},\ and\ \bibinfo
  {author} {\bibfnamefont {W.~D.}\ \bibnamefont {Oliver}},\ }\bibfield  {title}
  {\bibinfo {title} {Superconducting qubits: Current state of play},\ }\href
  {https://doi.org/https://doi.org/10.1146/annurev-conmatphys-031119-050605}
  {\bibfield  {journal} {\bibinfo  {journal} {Annual Review of Condensed Matter
  Physics}\ }\textbf {\bibinfo {volume} {11}},\ \bibinfo {pages} {369}
  (\bibinfo {year} {2020})}\BibitemShut {NoStop}%
\bibitem [{\citenamefont {Braum{\"u}ller}\ \emph {et~al.}(2017)\citenamefont
  {Braum{\"u}ller}, \citenamefont {Marthaler}, \citenamefont {Schneider},
  \citenamefont {Stehli}, \citenamefont {Rotzinger}, \citenamefont {Weides},\
  and\ \citenamefont {Ustinov}}]{Braumuller2017}%
  \BibitemOpen
  \bibfield  {author} {\bibinfo {author} {\bibfnamefont {J.}~\bibnamefont
  {Braum{\"u}ller}}, \bibinfo {author} {\bibfnamefont {M.}~\bibnamefont
  {Marthaler}}, \bibinfo {author} {\bibfnamefont {A.}~\bibnamefont
  {Schneider}}, \bibinfo {author} {\bibfnamefont {A.}~\bibnamefont {Stehli}},
  \bibinfo {author} {\bibfnamefont {H.}~\bibnamefont {Rotzinger}}, \bibinfo
  {author} {\bibfnamefont {M.}~\bibnamefont {Weides}},\ and\ \bibinfo {author}
  {\bibfnamefont {A.~V.}\ \bibnamefont {Ustinov}},\ }\bibfield  {title}
  {\bibinfo {title} {Analog quantum simulation of the rabi model in the
  ultra-strong coupling regime},\ }\href@noop {} {\bibfield  {journal}
  {\bibinfo  {journal} {Nature communications}\ }\textbf {\bibinfo {volume}
  {8}},\ \bibinfo {pages} {1} (\bibinfo {year} {2017})}\BibitemShut {NoStop}%
\bibitem [{\citenamefont {Ballester}\ \emph {et~al.}(2012)\citenamefont
  {Ballester}, \citenamefont {Romero}, \citenamefont {Garc\'{\i}a-Ripoll},
  \citenamefont {Deppe},\ and\ \citenamefont {Solano}}]{Ballester2012}%
  \BibitemOpen
  \bibfield  {author} {\bibinfo {author} {\bibfnamefont {D.}~\bibnamefont
  {Ballester}}, \bibinfo {author} {\bibfnamefont {G.}~\bibnamefont {Romero}},
  \bibinfo {author} {\bibfnamefont {J.~J.}\ \bibnamefont {Garc\'{\i}a-Ripoll}},
  \bibinfo {author} {\bibfnamefont {F.}~\bibnamefont {Deppe}},\ and\ \bibinfo
  {author} {\bibfnamefont {E.}~\bibnamefont {Solano}},\ }\bibfield  {title}
  {\bibinfo {title} {Quantum simulation of the ultrastrong-coupling dynamics in
  circuit quantum electrodynamics},\ }\href
  {https://doi.org/10.1103/PhysRevX.2.021007} {\bibfield  {journal} {\bibinfo
  {journal} {Phys. Rev. X}\ }\textbf {\bibinfo {volume} {2}},\ \bibinfo {pages}
  {021007} (\bibinfo {year} {2012})}\BibitemShut {NoStop}%
\bibitem [{\citenamefont {Wang}\ \emph {et~al.}(2020)\citenamefont {Wang},
  \citenamefont {Curtis}, \citenamefont {Lester}, \citenamefont {Zhang},
  \citenamefont {Gao}, \citenamefont {Freeze}, \citenamefont {Batista},
  \citenamefont {Vaccaro}, \citenamefont {Chuang}, \citenamefont {Frunzio},
  \citenamefont {Jiang}, \citenamefont {Girvin},\ and\ \citenamefont
  {Schoelkopf}}]{PhysRevX.10.021060}%
  \BibitemOpen
  \bibfield  {author} {\bibinfo {author} {\bibfnamefont {C.~S.}\ \bibnamefont
  {Wang}}, \bibinfo {author} {\bibfnamefont {J.~C.}\ \bibnamefont {Curtis}},
  \bibinfo {author} {\bibfnamefont {B.~J.}\ \bibnamefont {Lester}}, \bibinfo
  {author} {\bibfnamefont {Y.}~\bibnamefont {Zhang}}, \bibinfo {author}
  {\bibfnamefont {Y.~Y.}\ \bibnamefont {Gao}}, \bibinfo {author} {\bibfnamefont
  {J.}~\bibnamefont {Freeze}}, \bibinfo {author} {\bibfnamefont {V.~S.}\
  \bibnamefont {Batista}}, \bibinfo {author} {\bibfnamefont {P.~H.}\
  \bibnamefont {Vaccaro}}, \bibinfo {author} {\bibfnamefont {I.~L.}\
  \bibnamefont {Chuang}}, \bibinfo {author} {\bibfnamefont {L.}~\bibnamefont
  {Frunzio}}, \bibinfo {author} {\bibfnamefont {L.}~\bibnamefont {Jiang}},
  \bibinfo {author} {\bibfnamefont {S.~M.}\ \bibnamefont {Girvin}},\ and\
  \bibinfo {author} {\bibfnamefont {R.~J.}\ \bibnamefont {Schoelkopf}},\
  }\bibfield  {title} {\bibinfo {title} {Efficient multiphoton sampling of
  molecular vibronic spectra on a superconducting bosonic processor},\ }\href
  {https://doi.org/10.1103/PhysRevX.10.021060} {\bibfield  {journal} {\bibinfo
  {journal} {Phys. Rev. X}\ }\textbf {\bibinfo {volume} {10}},\ \bibinfo
  {pages} {021060} (\bibinfo {year} {2020})}\BibitemShut {NoStop}%
\bibitem [{\citenamefont {Flurin}\ \emph
  {et~al.}(2017{\natexlab{b}})\citenamefont {Flurin}, \citenamefont {Ramasesh},
  \citenamefont {Hacohen-Gourgy}, \citenamefont {Martin}, \citenamefont {Yao},\
  and\ \citenamefont {Siddiqi}}]{Flurin2017}%
  \BibitemOpen
  \bibfield  {author} {\bibinfo {author} {\bibfnamefont {E.}~\bibnamefont
  {Flurin}}, \bibinfo {author} {\bibfnamefont {V.~V.}\ \bibnamefont
  {Ramasesh}}, \bibinfo {author} {\bibfnamefont {S.}~\bibnamefont
  {Hacohen-Gourgy}}, \bibinfo {author} {\bibfnamefont {L.~S.}\ \bibnamefont
  {Martin}}, \bibinfo {author} {\bibfnamefont {N.~Y.}\ \bibnamefont {Yao}},\
  and\ \bibinfo {author} {\bibfnamefont {I.}~\bibnamefont {Siddiqi}},\
  }\bibfield  {title} {\bibinfo {title} {Observing topological invariants using
  quantum walks in superconducting circuits},\ }\href@noop {} {\bibfield
  {journal} {\bibinfo  {journal} {Physical Review X}\ }\textbf {\bibinfo
  {volume} {7}},\ \bibinfo {pages} {031023} (\bibinfo {year}
  {2017}{\natexlab{b}})}\BibitemShut {NoStop}%
\bibitem [{\citenamefont {Gerritsma}\ \emph {et~al.}(2010)\citenamefont
  {Gerritsma}, \citenamefont {Kirchmair}, \citenamefont {Z{\"a}hringer},
  \citenamefont {Solano}, \citenamefont {Blatt},\ and\ \citenamefont
  {Roos}}]{Gerritsma2010}%
  \BibitemOpen
  \bibfield  {author} {\bibinfo {author} {\bibfnamefont {R.}~\bibnamefont
  {Gerritsma}}, \bibinfo {author} {\bibfnamefont {G.}~\bibnamefont
  {Kirchmair}}, \bibinfo {author} {\bibfnamefont {F.}~\bibnamefont
  {Z{\"a}hringer}}, \bibinfo {author} {\bibfnamefont {E.}~\bibnamefont
  {Solano}}, \bibinfo {author} {\bibfnamefont {R.}~\bibnamefont {Blatt}},\ and\
  \bibinfo {author} {\bibfnamefont {C.}~\bibnamefont {Roos}},\ }\bibfield
  {title} {\bibinfo {title} {Quantum simulation of the dirac equation},\
  }\href@noop {} {\bibfield  {journal} {\bibinfo  {journal} {Nature}\ }\textbf
  {\bibinfo {volume} {463}},\ \bibinfo {pages} {68} (\bibinfo {year}
  {2010})}\BibitemShut {NoStop}%
\bibitem [{\citenamefont {Gerritsma}\ \emph {et~al.}(2011)\citenamefont
  {Gerritsma}, \citenamefont {Lanyon}, \citenamefont {Kirchmair}, \citenamefont
  {Z{\"a}hringer}, \citenamefont {Hempel}, \citenamefont {Casanova},
  \citenamefont {Garc{\'\i}a-Ripoll}, \citenamefont {Solano}, \citenamefont
  {Blatt},\ and\ \citenamefont {Roos}}]{Gerritsma2011}%
  \BibitemOpen
  \bibfield  {author} {\bibinfo {author} {\bibfnamefont {R.}~\bibnamefont
  {Gerritsma}}, \bibinfo {author} {\bibfnamefont {B.}~\bibnamefont {Lanyon}},
  \bibinfo {author} {\bibfnamefont {G.}~\bibnamefont {Kirchmair}}, \bibinfo
  {author} {\bibfnamefont {F.}~\bibnamefont {Z{\"a}hringer}}, \bibinfo {author}
  {\bibfnamefont {C.}~\bibnamefont {Hempel}}, \bibinfo {author} {\bibfnamefont
  {J.}~\bibnamefont {Casanova}}, \bibinfo {author} {\bibfnamefont {J.~J.}\
  \bibnamefont {Garc{\'\i}a-Ripoll}}, \bibinfo {author} {\bibfnamefont
  {E.}~\bibnamefont {Solano}}, \bibinfo {author} {\bibfnamefont
  {R.}~\bibnamefont {Blatt}},\ and\ \bibinfo {author} {\bibfnamefont {C.~F.}\
  \bibnamefont {Roos}},\ }\bibfield  {title} {\bibinfo {title} {Quantum
  simulation of the klein paradox with trapped ions},\ }\href@noop {}
  {\bibfield  {journal} {\bibinfo  {journal} {Physical review letters}\
  }\textbf {\bibinfo {volume} {106}},\ \bibinfo {pages} {060503} (\bibinfo
  {year} {2011})}\BibitemShut {NoStop}%
\bibitem [{\citenamefont {Casanova}\ \emph {et~al.}(2010)\citenamefont
  {Casanova}, \citenamefont {Garc\'{\i}a-Ripoll}, \citenamefont {Gerritsma},
  \citenamefont {Roos},\ and\ \citenamefont {Solano}}]{casanova2010}%
  \BibitemOpen
  \bibfield  {author} {\bibinfo {author} {\bibfnamefont {J.}~\bibnamefont
  {Casanova}}, \bibinfo {author} {\bibfnamefont {J.~J.}\ \bibnamefont
  {Garc\'{\i}a-Ripoll}}, \bibinfo {author} {\bibfnamefont {R.}~\bibnamefont
  {Gerritsma}}, \bibinfo {author} {\bibfnamefont {C.~F.}\ \bibnamefont
  {Roos}},\ and\ \bibinfo {author} {\bibfnamefont {E.}~\bibnamefont {Solano}},\
  }\bibfield  {title} {\bibinfo {title} {Klein tunneling and dirac potentials
  in trapped ions},\ }\href {https://doi.org/10.1103/PhysRevA.82.020101}
  {\bibfield  {journal} {\bibinfo  {journal} {Phys. Rev. A}\ }\textbf {\bibinfo
  {volume} {82}},\ \bibinfo {pages} {020101} (\bibinfo {year}
  {2010})}\BibitemShut {NoStop}%
\bibitem [{\citenamefont {Jiang}\ \emph {et~al.}(2022)\citenamefont {Jiang},
  \citenamefont {Cai}, \citenamefont {Wu}, \citenamefont {Mei}, \citenamefont
  {Zhao}, \citenamefont {Chang}, \citenamefont {Yao}, \citenamefont {He},
  \citenamefont {Zhou},\ and\ \citenamefont {Duan}}]{Jiang2022}%
  \BibitemOpen
  \bibfield  {author} {\bibinfo {author} {\bibfnamefont {Y.}~\bibnamefont
  {Jiang}}, \bibinfo {author} {\bibfnamefont {M.-L.}\ \bibnamefont {Cai}},
  \bibinfo {author} {\bibfnamefont {Y.-K.}\ \bibnamefont {Wu}}, \bibinfo
  {author} {\bibfnamefont {Q.-X.}\ \bibnamefont {Mei}}, \bibinfo {author}
  {\bibfnamefont {W.-D.}\ \bibnamefont {Zhao}}, \bibinfo {author}
  {\bibfnamefont {X.-Y.}\ \bibnamefont {Chang}}, \bibinfo {author}
  {\bibfnamefont {L.}~\bibnamefont {Yao}}, \bibinfo {author} {\bibfnamefont
  {L.}~\bibnamefont {He}}, \bibinfo {author} {\bibfnamefont {Z.-C.}\
  \bibnamefont {Zhou}},\ and\ \bibinfo {author} {\bibfnamefont {L.-M.}\
  \bibnamefont {Duan}},\ }\bibfield  {title} {\bibinfo {title} {Quantum
  simulation of the two-dimensional weyl equation in a magnetic field},\ }\href
  {https://doi.org/10.1103/PhysRevLett.128.200502} {\bibfield  {journal}
  {\bibinfo  {journal} {Phys. Rev. Lett.}\ }\textbf {\bibinfo {volume} {128}},\
  \bibinfo {pages} {200502} (\bibinfo {year} {2022})}\BibitemShut {NoStop}%
\bibitem [{\citenamefont {Svetitsky}\ and\ \citenamefont
  {Katz}(2019)}]{Svetitsky2019}%
  \BibitemOpen
  \bibfield  {author} {\bibinfo {author} {\bibfnamefont {E.}~\bibnamefont
  {Svetitsky}}\ and\ \bibinfo {author} {\bibfnamefont {N.}~\bibnamefont
  {Katz}},\ }\bibfield  {title} {\bibinfo {title} {Dirac particle dynamics of a
  superconducting circuit},\ }\href@noop {} {\bibfield  {journal} {\bibinfo
  {journal} {Physical Review A}\ }\textbf {\bibinfo {volume} {99}},\ \bibinfo
  {pages} {042308} (\bibinfo {year} {2019})}\BibitemShut {NoStop}%
\bibitem [{\citenamefont {Pedernales}\ \emph {et~al.}(2013)\citenamefont
  {Pedernales}, \citenamefont {Di~Candia}, \citenamefont {Ballester},\ and\
  \citenamefont {Solano}}]{Pedernales_2013}%
  \BibitemOpen
  \bibfield  {author} {\bibinfo {author} {\bibfnamefont {J.~S.}\ \bibnamefont
  {Pedernales}}, \bibinfo {author} {\bibfnamefont {R.}~\bibnamefont
  {Di~Candia}}, \bibinfo {author} {\bibfnamefont {D.}~\bibnamefont
  {Ballester}},\ and\ \bibinfo {author} {\bibfnamefont {E.}~\bibnamefont
  {Solano}},\ }\bibfield  {title} {\bibinfo {title} {Quantum simulations of
  relativistic quantum physics in circuit qed},\ }\href
  {https://doi.org/10.1088/1367-2630/15/5/055008} {\bibfield  {journal}
  {\bibinfo  {journal} {New Journal of Physics}\ }\textbf {\bibinfo {volume}
  {15}},\ \bibinfo {pages} {055008} (\bibinfo {year} {2013})}\BibitemShut
  {NoStop}%
\bibitem [{\citenamefont {Klein}(1929)}]{Klein1929}%
  \BibitemOpen
  \bibfield  {author} {\bibinfo {author} {\bibfnamefont {O.}~\bibnamefont
  {Klein}},\ }\bibfield  {title} {\bibinfo {title} {Die reflexion von
  elektronen an einem potentialsprung nach der relativistischen dynamik von
  dirac},\ }\href@noop {} {\bibfield  {journal} {\bibinfo  {journal}
  {Zeitschrift f{\"u}r Physik}\ }\textbf {\bibinfo {volume} {53}},\ \bibinfo
  {pages} {157} (\bibinfo {year} {1929})}\BibitemShut {NoStop}%
\bibitem [{\citenamefont {Thaller}(2013)}]{Thaller2013}%
  \BibitemOpen
  \bibfield  {author} {\bibinfo {author} {\bibfnamefont {B.}~\bibnamefont
  {Thaller}},\ }\href@noop {} {\emph {\bibinfo {title} {The dirac equation}}}\
  (\bibinfo  {publisher} {Springer Science \& Business Media},\ \bibinfo {year}
  {2013})\BibitemShut {NoStop}%
\bibitem [{\citenamefont {Anderson}(1933)}]{Anderson1933}%
  \BibitemOpen
  \bibfield  {author} {\bibinfo {author} {\bibfnamefont {C.~D.}\ \bibnamefont
  {Anderson}},\ }\bibfield  {title} {\bibinfo {title} {The positive electron},\
  }\href@noop {} {\bibfield  {journal} {\bibinfo  {journal} {Physical Review}\
  }\textbf {\bibinfo {volume} {43}},\ \bibinfo {pages} {491} (\bibinfo {year}
  {1933})}\BibitemShut {NoStop}%
\bibitem [{\citenamefont {Thaller}(2004)}]{thaller2004}%
  \BibitemOpen
  \bibfield  {author} {\bibinfo {author} {\bibfnamefont {B.}~\bibnamefont
  {Thaller}},\ }\href {https://arxiv.org/abs/quant-ph/0409079} {\bibinfo
  {title} {Visualizing the kinematics of relativistic wave packets}} (\bibinfo
  {year} {2004}),\ \Eprint {https://arxiv.org/abs/quant-ph/0409079}
  {arXiv:quant-ph/0409079 [quant-ph]} \BibitemShut {NoStop}%
\bibitem [{\citenamefont {Hacohen-Gourgy}\ \emph {et~al.}(2016)\citenamefont
  {Hacohen-Gourgy}, \citenamefont {Martin}, \citenamefont {Flurin},
  \citenamefont {Ramasesh}, \citenamefont {Whaley},\ and\ \citenamefont
  {Siddiqi}}]{Hacohen2016}%
  \BibitemOpen
  \bibfield  {author} {\bibinfo {author} {\bibfnamefont {S.}~\bibnamefont
  {Hacohen-Gourgy}}, \bibinfo {author} {\bibfnamefont {L.~S.}\ \bibnamefont
  {Martin}}, \bibinfo {author} {\bibfnamefont {E.}~\bibnamefont {Flurin}},
  \bibinfo {author} {\bibfnamefont {V.~V.}\ \bibnamefont {Ramasesh}}, \bibinfo
  {author} {\bibfnamefont {K.~B.}\ \bibnamefont {Whaley}},\ and\ \bibinfo
  {author} {\bibfnamefont {I.}~\bibnamefont {Siddiqi}},\ }\bibfield  {title}
  {\bibinfo {title} {Quantum dynamics of simultaneously measured non-commuting
  observables},\ }\href@noop {} {\bibfield  {journal} {\bibinfo  {journal}
  {Nature}\ }\textbf {\bibinfo {volume} {538}},\ \bibinfo {pages} {491}
  (\bibinfo {year} {2016})}\BibitemShut {NoStop}%
\bibitem [{\citenamefont {Murch}\ \emph {et~al.}(2012)\citenamefont {Murch},
  \citenamefont {Vool}, \citenamefont {Zhou}, \citenamefont {Weber},
  \citenamefont {Girvin},\ and\ \citenamefont
  {Siddiqi}}]{PhysRevLett.109.183602}%
  \BibitemOpen
  \bibfield  {author} {\bibinfo {author} {\bibfnamefont {K.~W.}\ \bibnamefont
  {Murch}}, \bibinfo {author} {\bibfnamefont {U.}~\bibnamefont {Vool}},
  \bibinfo {author} {\bibfnamefont {D.}~\bibnamefont {Zhou}}, \bibinfo {author}
  {\bibfnamefont {S.~J.}\ \bibnamefont {Weber}}, \bibinfo {author}
  {\bibfnamefont {S.~M.}\ \bibnamefont {Girvin}},\ and\ \bibinfo {author}
  {\bibfnamefont {I.}~\bibnamefont {Siddiqi}},\ }\bibfield  {title} {\bibinfo
  {title} {Cavity-assisted quantum bath engineering},\ }\href
  {https://doi.org/10.1103/PhysRevLett.109.183602} {\bibfield  {journal}
  {\bibinfo  {journal} {Phys. Rev. Lett.}\ }\textbf {\bibinfo {volume} {109}},\
  \bibinfo {pages} {183602} (\bibinfo {year} {2012})}\BibitemShut {NoStop}%
\bibitem [{\citenamefont {Diringer}\ \emph {et~al.}(2024)\citenamefont
  {Diringer}, \citenamefont {Blumenthal}, \citenamefont {Grinberg},
  \citenamefont {Jiang},\ and\ \citenamefont
  {Hacohen-Gourgy}}]{PhysRevX.14.011055}%
  \BibitemOpen
  \bibfield  {author} {\bibinfo {author} {\bibfnamefont {A.~A.}\ \bibnamefont
  {Diringer}}, \bibinfo {author} {\bibfnamefont {E.}~\bibnamefont
  {Blumenthal}}, \bibinfo {author} {\bibfnamefont {A.}~\bibnamefont
  {Grinberg}}, \bibinfo {author} {\bibfnamefont {L.}~\bibnamefont {Jiang}},\
  and\ \bibinfo {author} {\bibfnamefont {S.}~\bibnamefont {Hacohen-Gourgy}},\
  }\bibfield  {title} {\bibinfo {title} {Conditional-not displacement: Fast
  multioscillator control with a single qubit},\ }\href
  {https://doi.org/10.1103/PhysRevX.14.011055} {\bibfield  {journal} {\bibinfo
  {journal} {Phys. Rev. X}\ }\textbf {\bibinfo {volume} {14}},\ \bibinfo
  {pages} {011055} (\bibinfo {year} {2024})}\BibitemShut {NoStop}%
\bibitem [{\citenamefont {Magnus}(1954)}]{Magnus1954}%
  \BibitemOpen
  \bibfield  {author} {\bibinfo {author} {\bibfnamefont {W.}~\bibnamefont
  {Magnus}},\ }\bibfield  {title} {\bibinfo {title} {On the exponential
  solution of differential equations for a linear operator},\ }\href
  {https://doi.org/https://doi.org/10.1002/cpa.3160070404} {\bibfield
  {journal} {\bibinfo  {journal} {Communications on Pure and Applied
  Mathematics}\ }\textbf {\bibinfo {volume} {7}},\ \bibinfo {pages} {649}
  (\bibinfo {year} {1954})},\ \Eprint
  {https://arxiv.org/abs/https://onlinelibrary.wiley.com/doi/pdf/10.1002/cpa.3160070404}
  {https://onlinelibrary.wiley.com/doi/pdf/10.1002/cpa.3160070404} \BibitemShut
  {NoStop}%
\bibitem [{\citenamefont {{Landau}}(1930)}]{Landau1930}%
  \BibitemOpen
  \bibfield  {author} {\bibinfo {author} {\bibfnamefont {L.}~\bibnamefont
  {{Landau}}},\ }\bibfield  {title} {\bibinfo {title} {{Diamagnetismus der
  Metalle}},\ }\href {https://doi.org/10.1007/BF01397213} {\bibfield  {journal}
  {\bibinfo  {journal} {Zeitschrift fur Physik}\ }\textbf {\bibinfo {volume}
  {64}},\ \bibinfo {pages} {629} (\bibinfo {year} {1930})}\BibitemShut
  {NoStop}%
\bibitem [{\citenamefont {Lamata}\ \emph {et~al.}(2011)\citenamefont {Lamata},
  \citenamefont {Casanova}, \citenamefont {Gerritsma}, \citenamefont {Roos},
  \citenamefont {García-Ripoll},\ and\ \citenamefont {Solano}}]{Lamata2011}%
  \BibitemOpen
  \bibfield  {author} {\bibinfo {author} {\bibfnamefont {L.}~\bibnamefont
  {Lamata}}, \bibinfo {author} {\bibfnamefont {J.}~\bibnamefont {Casanova}},
  \bibinfo {author} {\bibfnamefont {R.}~\bibnamefont {Gerritsma}}, \bibinfo
  {author} {\bibfnamefont {C.~F.}\ \bibnamefont {Roos}}, \bibinfo {author}
  {\bibfnamefont {J.~J.}\ \bibnamefont {García-Ripoll}},\ and\ \bibinfo
  {author} {\bibfnamefont {E.}~\bibnamefont {Solano}},\ }\bibfield  {title}
  {\bibinfo {title} {Relativistic quantum mechanics with trapped ions},\ }\href
  {https://doi.org/10.1088/1367-2630/13/9/095003} {\bibfield  {journal}
  {\bibinfo  {journal} {New Journal of Physics}\ }\textbf {\bibinfo {volume}
  {13}},\ \bibinfo {pages} {095003} (\bibinfo {year} {2011})}\BibitemShut
  {NoStop}%
\bibitem [{\citenamefont {Schwinger}(1951)}]{PhysRev.82.664}%
  \BibitemOpen
  \bibfield  {author} {\bibinfo {author} {\bibfnamefont {J.}~\bibnamefont
  {Schwinger}},\ }\bibfield  {title} {\bibinfo {title} {On gauge invariance and
  vacuum polarization},\ }\href {https://doi.org/10.1103/PhysRev.82.664}
  {\bibfield  {journal} {\bibinfo  {journal} {Phys. Rev.}\ }\textbf {\bibinfo
  {volume} {82}},\ \bibinfo {pages} {664} (\bibinfo {year} {1951})}\BibitemShut
  {NoStop}%
\bibitem [{\citenamefont {Hund}(1941)}]{Hund1941}%
  \BibitemOpen
  \bibfield  {author} {\bibinfo {author} {\bibfnamefont {F.}~\bibnamefont
  {Hund}},\ }\bibfield  {title} {\bibinfo {title} {Materieerzeugung im
  anschaulichen und im gequantelten wellenbild der materie},\ }\href
  {https://doi.org/10.1007/BF01337403} {\bibfield  {journal} {\bibinfo
  {journal} {Zeitschrift f{\"u}r Physik}\ }\textbf {\bibinfo {volume} {117}},\
  \bibinfo {pages} {1} (\bibinfo {year} {1941})}\BibitemShut {NoStop}%
\bibitem [{\citenamefont {Rubbmark}\ \emph {et~al.}(1981)\citenamefont
  {Rubbmark}, \citenamefont {Kash}, \citenamefont {Littman},\ and\
  \citenamefont {Kleppner}}]{PhysRevA.23.3107}%
  \BibitemOpen
  \bibfield  {author} {\bibinfo {author} {\bibfnamefont {J.~R.}\ \bibnamefont
  {Rubbmark}}, \bibinfo {author} {\bibfnamefont {M.~M.}\ \bibnamefont {Kash}},
  \bibinfo {author} {\bibfnamefont {M.~G.}\ \bibnamefont {Littman}},\ and\
  \bibinfo {author} {\bibfnamefont {D.}~\bibnamefont {Kleppner}},\ }\bibfield
  {title} {\bibinfo {title} {Dynamical effects at avoided level crossings: A
  study of the landau-zener effect using rydberg atoms},\ }\href
  {https://doi.org/10.1103/PhysRevA.23.3107} {\bibfield  {journal} {\bibinfo
  {journal} {Phys. Rev. A}\ }\textbf {\bibinfo {volume} {23}},\ \bibinfo
  {pages} {3107} (\bibinfo {year} {1981})}\BibitemShut {NoStop}%
\bibitem [{\citenamefont {Chakram}\ \emph {et~al.}(2022)\citenamefont
  {Chakram}, \citenamefont {He}, \citenamefont {Dixit}, \citenamefont {Oriani},
  \citenamefont {Naik}, \citenamefont {Leung}, \citenamefont {Kwon},
  \citenamefont {Ma}, \citenamefont {Jiang},\ and\ \citenamefont
  {Schuster}}]{chakram2022multimode}%
  \BibitemOpen
  \bibfield  {author} {\bibinfo {author} {\bibfnamefont {S.}~\bibnamefont
  {Chakram}}, \bibinfo {author} {\bibfnamefont {K.}~\bibnamefont {He}},
  \bibinfo {author} {\bibfnamefont {A.~V.}\ \bibnamefont {Dixit}}, \bibinfo
  {author} {\bibfnamefont {A.~E.}\ \bibnamefont {Oriani}}, \bibinfo {author}
  {\bibfnamefont {R.~K.}\ \bibnamefont {Naik}}, \bibinfo {author}
  {\bibfnamefont {N.}~\bibnamefont {Leung}}, \bibinfo {author} {\bibfnamefont
  {H.}~\bibnamefont {Kwon}}, \bibinfo {author} {\bibfnamefont {W.-L.}\
  \bibnamefont {Ma}}, \bibinfo {author} {\bibfnamefont {L.}~\bibnamefont
  {Jiang}},\ and\ \bibinfo {author} {\bibfnamefont {D.~I.}\ \bibnamefont
  {Schuster}},\ }\bibfield  {title} {\bibinfo {title} {Multimode photon
  blockade},\ }\href@noop {} {\bibfield  {journal} {\bibinfo  {journal} {Nature
  Physics}\ }\textbf {\bibinfo {volume} {18}},\ \bibinfo {pages} {879}
  (\bibinfo {year} {2022})}\BibitemShut {NoStop}%
\bibitem [{\citenamefont {Lu}\ \emph {et~al.}(2022)\citenamefont {Lu},
  \citenamefont {Ville}, \citenamefont {Cohen}, \citenamefont {Petrescu},
  \citenamefont {Schreppler}, \citenamefont {Chen}, \citenamefont {J\"unger},
  \citenamefont {Pelletti}, \citenamefont {Marchenkov}, \citenamefont
  {Banerjee}, \citenamefont {Livingston}, \citenamefont {Kreikebaum},
  \citenamefont {Santiago}, \citenamefont {Blais},\ and\ \citenamefont
  {Siddiqi}}]{Lu2022}%
  \BibitemOpen
  \bibfield  {author} {\bibinfo {author} {\bibfnamefont {M.}~\bibnamefont
  {Lu}}, \bibinfo {author} {\bibfnamefont {J.-L.}\ \bibnamefont {Ville}},
  \bibinfo {author} {\bibfnamefont {J.}~\bibnamefont {Cohen}}, \bibinfo
  {author} {\bibfnamefont {A.}~\bibnamefont {Petrescu}}, \bibinfo {author}
  {\bibfnamefont {S.}~\bibnamefont {Schreppler}}, \bibinfo {author}
  {\bibfnamefont {L.}~\bibnamefont {Chen}}, \bibinfo {author} {\bibfnamefont
  {C.}~\bibnamefont {J\"unger}}, \bibinfo {author} {\bibfnamefont
  {C.}~\bibnamefont {Pelletti}}, \bibinfo {author} {\bibfnamefont
  {A.}~\bibnamefont {Marchenkov}}, \bibinfo {author} {\bibfnamefont
  {A.}~\bibnamefont {Banerjee}}, \bibinfo {author} {\bibfnamefont {W.~P.}\
  \bibnamefont {Livingston}}, \bibinfo {author} {\bibfnamefont {J.~M.}\
  \bibnamefont {Kreikebaum}}, \bibinfo {author} {\bibfnamefont {D.~I.}\
  \bibnamefont {Santiago}}, \bibinfo {author} {\bibfnamefont {A.}~\bibnamefont
  {Blais}},\ and\ \bibinfo {author} {\bibfnamefont {I.}~\bibnamefont
  {Siddiqi}},\ }\bibfield  {title} {\bibinfo {title} {Multipartite entanglement
  in rabi-driven superconducting qubits},\ }\href
  {https://doi.org/10.1103/PRXQuantum.3.040322} {\bibfield  {journal} {\bibinfo
   {journal} {PRX Quantum}\ }\textbf {\bibinfo {volume} {3}},\ \bibinfo {pages}
  {040322} (\bibinfo {year} {2022})}\BibitemShut {NoStop}%
\bibitem [{\citenamefont {Milul}\ \emph {et~al.}(2023)\citenamefont {Milul},
  \citenamefont {Guttel}, \citenamefont {Goldblatt}, \citenamefont {Hazanov},
  \citenamefont {Joshi}, \citenamefont {Chausovsky}, \citenamefont {Kahn},
  \citenamefont {{\c{C}}ifty{\"u}rek}, \citenamefont {Lafont},\ and\
  \citenamefont {Rosenblum}}]{milul2023superconducting}%
  \BibitemOpen
  \bibfield  {author} {\bibinfo {author} {\bibfnamefont {O.}~\bibnamefont
  {Milul}}, \bibinfo {author} {\bibfnamefont {B.}~\bibnamefont {Guttel}},
  \bibinfo {author} {\bibfnamefont {U.}~\bibnamefont {Goldblatt}}, \bibinfo
  {author} {\bibfnamefont {S.}~\bibnamefont {Hazanov}}, \bibinfo {author}
  {\bibfnamefont {L.~M.}\ \bibnamefont {Joshi}}, \bibinfo {author}
  {\bibfnamefont {D.}~\bibnamefont {Chausovsky}}, \bibinfo {author}
  {\bibfnamefont {N.}~\bibnamefont {Kahn}}, \bibinfo {author} {\bibfnamefont
  {E.}~\bibnamefont {{\c{C}}ifty{\"u}rek}}, \bibinfo {author} {\bibfnamefont
  {F.}~\bibnamefont {Lafont}},\ and\ \bibinfo {author} {\bibfnamefont
  {S.}~\bibnamefont {Rosenblum}},\ }\bibfield  {title} {\bibinfo {title}
  {Superconducting cavity qubit with tens of milliseconds single-photon
  coherence time},\ }\href@noop {} {\bibfield  {journal} {\bibinfo  {journal}
  {PRX Quantum}\ }\textbf {\bibinfo {volume} {4}},\ \bibinfo {pages} {030336}
  (\bibinfo {year} {2023})}\BibitemShut {NoStop}%
\bibitem [{\citenamefont {Tuokkola}\ \emph {et~al.}(2024)\citenamefont
  {Tuokkola}, \citenamefont {Sunada}, \citenamefont {Kivijärvi}, \citenamefont
  {Albanese}, \citenamefont {Grönberg}, \citenamefont {Kaikkonen},
  \citenamefont {Vesterinen}, \citenamefont {Govenius},\ and\ \citenamefont
  {Möttönen}}]{tuokkola2024}%
  \BibitemOpen
  \bibfield  {author} {\bibinfo {author} {\bibfnamefont {M.}~\bibnamefont
  {Tuokkola}}, \bibinfo {author} {\bibfnamefont {Y.}~\bibnamefont {Sunada}},
  \bibinfo {author} {\bibfnamefont {H.}~\bibnamefont {Kivijärvi}}, \bibinfo
  {author} {\bibfnamefont {J.}~\bibnamefont {Albanese}}, \bibinfo {author}
  {\bibfnamefont {L.}~\bibnamefont {Grönberg}}, \bibinfo {author}
  {\bibfnamefont {J.-P.}\ \bibnamefont {Kaikkonen}}, \bibinfo {author}
  {\bibfnamefont {V.}~\bibnamefont {Vesterinen}}, \bibinfo {author}
  {\bibfnamefont {J.}~\bibnamefont {Govenius}},\ and\ \bibinfo {author}
  {\bibfnamefont {M.}~\bibnamefont {Möttönen}},\ }\href
  {https://arxiv.org/abs/2407.18778} {\bibinfo {title} {Methods to achieve
  near-millisecond energy relaxation and dephasing times for a superconducting
  transmon qubit}} (\bibinfo {year} {2024}),\ \Eprint
  {https://arxiv.org/abs/2407.18778} {arXiv:2407.18778 [quant-ph]} \BibitemShut
  {NoStop}%
\bibitem [{\citenamefont {Novoselov}\ \emph {et~al.}(2005)\citenamefont
  {Novoselov}, \citenamefont {Geim}, \citenamefont {Morozov}, \citenamefont
  {Jiang}, \citenamefont {Katsnelson}, \citenamefont {Grigorieva},
  \citenamefont {Dubonos},\ and\ \citenamefont {Firsov}}]{Novoselov2005}%
  \BibitemOpen
  \bibfield  {author} {\bibinfo {author} {\bibfnamefont {K.~S.}\ \bibnamefont
  {Novoselov}}, \bibinfo {author} {\bibfnamefont {A.~K.}\ \bibnamefont {Geim}},
  \bibinfo {author} {\bibfnamefont {S.~V.}\ \bibnamefont {Morozov}}, \bibinfo
  {author} {\bibfnamefont {D.}~\bibnamefont {Jiang}}, \bibinfo {author}
  {\bibfnamefont {M.~I.}\ \bibnamefont {Katsnelson}}, \bibinfo {author}
  {\bibfnamefont {I.~V.}\ \bibnamefont {Grigorieva}}, \bibinfo {author}
  {\bibfnamefont {S.~V.}\ \bibnamefont {Dubonos}},\ and\ \bibinfo {author}
  {\bibfnamefont {A.~A.}\ \bibnamefont {Firsov}},\ }\bibfield  {title}
  {\bibinfo {title} {Two-dimensional gas of massless dirac fermions in
  graphene},\ }\href {https://doi.org/10.1038/nature04233} {\bibfield
  {journal} {\bibinfo  {journal} {Nature}\ }\textbf {\bibinfo {volume} {438}},\
  \bibinfo {pages} {197} (\bibinfo {year} {2005})}\BibitemShut {NoStop}%
\bibitem [{\citenamefont {Lamata}(2017)}]{Lamata2017}%
  \BibitemOpen
  \bibfield  {author} {\bibinfo {author} {\bibfnamefont {L.}~\bibnamefont
  {Lamata}},\ }\bibfield  {title} {\bibinfo {title} {Digital-analog quantum
  simulation of generalized dicke models with superconducting circuits},\
  }\href {https://doi.org/10.1038/srep43768} {\bibfield  {journal} {\bibinfo
  {journal} {Scientific Reports}\ }\textbf {\bibinfo {volume} {7}},\ \bibinfo
  {pages} {43768} (\bibinfo {year} {2017})}\BibitemShut {NoStop}%
\bibitem [{\citenamefont {Mezzacapo}\ \emph {et~al.}(2014)\citenamefont
  {Mezzacapo}, \citenamefont {Las~Heras}, \citenamefont {Pedernales},
  \citenamefont {DiCarlo}, \citenamefont {Solano},\ and\ \citenamefont
  {Lamata}}]{Mezzacapo2014}%
  \BibitemOpen
  \bibfield  {author} {\bibinfo {author} {\bibfnamefont {A.}~\bibnamefont
  {Mezzacapo}}, \bibinfo {author} {\bibfnamefont {U.}~\bibnamefont
  {Las~Heras}}, \bibinfo {author} {\bibfnamefont {J.~S.}\ \bibnamefont
  {Pedernales}}, \bibinfo {author} {\bibfnamefont {L.}~\bibnamefont {DiCarlo}},
  \bibinfo {author} {\bibfnamefont {E.}~\bibnamefont {Solano}},\ and\ \bibinfo
  {author} {\bibfnamefont {L.}~\bibnamefont {Lamata}},\ }\bibfield  {title}
  {\bibinfo {title} {Digital quantum rabi and dicke models in superconducting
  circuits},\ }\href {https://doi.org/10.1038/srep07482} {\bibfield  {journal}
  {\bibinfo  {journal} {Scientific Reports}\ }\textbf {\bibinfo {volume} {4}},\
  \bibinfo {pages} {7482} (\bibinfo {year} {2014})}\BibitemShut {NoStop}%
\end{thebibliography}%

\end{document}